\documentclass[secnumarabic,amssymb,pra,reprint,superscriptaddress,longbibliography]{revtex4-2}
\usepackage[english]{babel}
\usepackage{graphicx}
\usepackage{amsmath}
\usepackage{amsfonts}
\usepackage[usenames,dvipsnames]{color}
\usepackage{hyperref}
\usepackage{blindtext}
\usepackage{soul}
\hypersetup{colorlinks}
\usepackage[utf8]{inputenc}
\usepackage{color}
\usepackage[version=3]{mhchem} 
\usepackage{ifthen}
\hypersetup{linkcolor=black, citecolor=blue, urlcolor=blue}

\makeatletter
\addto\captionsenglish{\renewcommand{\figurename}{Fig.}}
\renewcommand*{\fnum@figure}{{\normalfont\bfseries \figurename~\thefigure}}
\renewcommand*{\@caption@fignum@sep}{\textbf{ }}
\makeatother
\setcitestyle{super}

\begin{document}
\title{Anisotropic electrostatic screening of charged colloids in nematic solvents}

\author{Jeffrey C. Everts}
\thanks{These two authors contributed equally}
\affiliation{Faculty of Mathematics and Physics, University of Ljubljana, Jadranska 19, 1000 Ljubljana, Slovenia.}
\affiliation{Institute of Physical Chemistry, Polish Academy of Sciences, Kasprzaka 44/52, PL-01-224 Warsaw, Poland}
\author{Bohdan Senyuk}
\thanks{These authors contributed equally to this work.}
\affiliation{Department of Physics and Soft Materials Research Center, University of Colorado, Boulder, CO 80309, USA.}

\author{Haridas Mundoor}
\affiliation{Department of Physics and Soft Materials Research Center, University of Colorado, Boulder, CO 80309, USA.}

\author{Miha Ravnik}
\email{miha.ravnik@fmf.uni-lj.si}
\affiliation{Faculty of Mathematics and Physics, University of Ljubljana, Jadranska 19, 1000 Ljubljana, Slovenia.}
\affiliation{Jozef Stefan Institute, Jamova 39, 1000 Ljubljana, Slovenia.}

\author{Ivan I. Smalyukh}
\email{ivan.smalyukh@colorado.edu}
\affiliation{Department of Physics and Soft Materials Research Center, University of Colorado, Boulder, CO 80309, USA.}
\affiliation{Department of Electrical, Computer and Energy Engineering and Materials Science and Engineering Program, University of Colorado, Boulder, CO 80309, USA.}
\affiliation{Renewable and Sustainable Energy Institute, National Renewable Energy Laboratory, University of Colorado, Boulder, CO 80309, USA.
}

\date{\today}

\begin{abstract}
{The physical behaviour of anisotropic charged colloids is determined by their material dielectric anisotropy, affecting colloidal
self-assembly, biological function and even out-of-equilibrium behaviour. However, little is known about
anisotropic electrostatic screening, which underlies all electrostatic effective interactions in such soft or
biological materials. In this work, we demonstrate anisotropic electrostatic screening for charged colloidal particles in a nematic electrolyte. We show that material anisotropy
behaves markedly different from particle anisotropy: The electrostatic potential and pair interactions
decay with an anisotropic Debye screening length, contrasting the constant screening length for isotropic electrolytes. Charged dumpling-shaped near-spherical colloidal particles in a nematic medium are used as an experimental model system to explore the effects of anisotropic screening,
demonstrating competing anisotropic elastic and electrostatic effective pair interactions for colloidal
surface charges tunable from neutral to high, yielding particle-separated metastable states. Generally,
our work contributes to the understanding of electrostatic screening in nematic anisotropic media.
}
\end{abstract}
\maketitle
\section*{Introduction}

Highly charged biopolymers like DNA and filamentous actin are just two of many examples of biological relevance of electrostatic interactions that are screened by counterions under physiological conditions. In soft condensed matter, similar effects allow for exploiting electrostatic interactions between particles in defining colloidal self-organized superstructures that they can form and, even more importantly, enabling the very existence of metastable colloidal systems. \textcolor{black}{These structures range from regular --- linear, two-dimensional (2D), and 3D  crystalline --- structures to amorphous structures and have broad relevance, including in photonics, optics, paint, and food industry.} Screened electrostatic interactions have been studied quite extensively in systems where the solvent is isotropic. Findings include colloid stabilisation by charge \cite{Derjaguin:1948, VerweyOverbeek}, the description of short-range liquid order in scattering experiments \cite{Veretout:1989}, measurements and calculations of pair interactions between two charged-screened particles \cite{Ducker:1991, Carnie:1993, Grier:1994, Borkovec:2018}, ion transport \cite{Zhang:2019}, the influence\textcolor{black} of external fields \cite{Lowen:2002, Dijkstra:2005, Zaccone:2009} and predicting phase behaviour, such as demixing \cite{Evans:2006, Royall:2012} and crystallisation \cite{Leunissen:2005, Li:2017}. Furthermore, there are many theoretical studies on many-body effects \cite{Roij:1997, Levin:2004, Everts:2016}, charge regulation \cite{Ninham:1971, Borkovec:2010, Levin:2019}, charge renormalisation \cite{Alexander:1984}, charge fluctuations \cite{Naji:2013}, and nonadditive effects of dispersion interactions \cite{Ninham:2001, Kotov:2015}. Despite these extensive studies for particles dispersed in isotropic electrolytes, the understanding of electrostatic screening in ion-doped nematic media seems to be lacking. \textcolor{black}{Anisotropic electrostatic screening can potentially provide the means for controlling and engineering self-assembly of colloidal particles in nematic solvents, where a key advantage as compared with isotropic solvents may arise in defining highly anisotropic interactions and self-assembled structures. An important relevance of anisotropic electrostatic screening is also in active matter systems, as affecting major mechanisms including locomotion and energy harvesting \cite{Gompper:2020} } Here, we address the issue of electrostatic screening in nematic media and show that it is important by exploring an experimental model system in media with orientational elasticity.

{\color{black}
For isotropic solvents and sufficiently dilute electrolytes, the electrostatic potential around a freely-dispersed arbitrarily-shaped charged particle asymptotically scales as \cite{Trizac:2000, Trizac:2002, Tellez:2010}
\begin{equation}
\phi({\bf r})\sim\mathcal{A}\left(\psi,\theta;\lambda_\mathrm{D}^\mathrm{I}\right)\frac{\exp\left(-r/\lambda_\mathrm{D}^\mathrm{I}\right)}{r}, \quad (r\rightarrow\infty),
\label{eq:anisop}
\end{equation}
with $r$ as the radial distance, $\psi$ as the azimuthal angle, $\theta$ as the polar angle, and $\lambda_\mathrm{D}^\mathrm{I}$ as the constant (isotropic) Debye screening length. The anisotropy function $\mathcal{A}(\psi,\theta;\lambda_\mathrm{D}^\mathrm{I})$ captures the shape effects of the particle on $\phi({\bf r})$, and is independent of $\psi$ and $\theta$ for spherical particles at any value of $\lambda_\mathrm{D}^\mathrm{I}$. In general $\mathcal{A}$ becomes more strongly dependent on $(\psi,\theta)$ when $\lambda_\mathrm{D}^\mathrm{I}$ is small, whereas $\mathcal{A}(\psi,\theta;\lambda_\mathrm{D}^\mathrm{I})\rightarrow 1$ for $\lambda_\mathrm{D}^\mathrm{I}\rightarrow\infty$, meaning that $\phi({\bf r})\sim 1/r$ for \emph{any} particle. In other words, anisotropies stemming from the particle shape are present even asymptotically far from the particle, contrasting the unscreened case that behaves as a point charge for $r\rightarrow\infty$. Furthermore, note that dilute isotropic electrolytes are characterised by a single value of  $\lambda^\mathrm{I}_\mathrm{D}$ independent of the particle orientation.}

\begin{figure*}[t]
\centering
\includegraphics[width=0.9\textwidth]{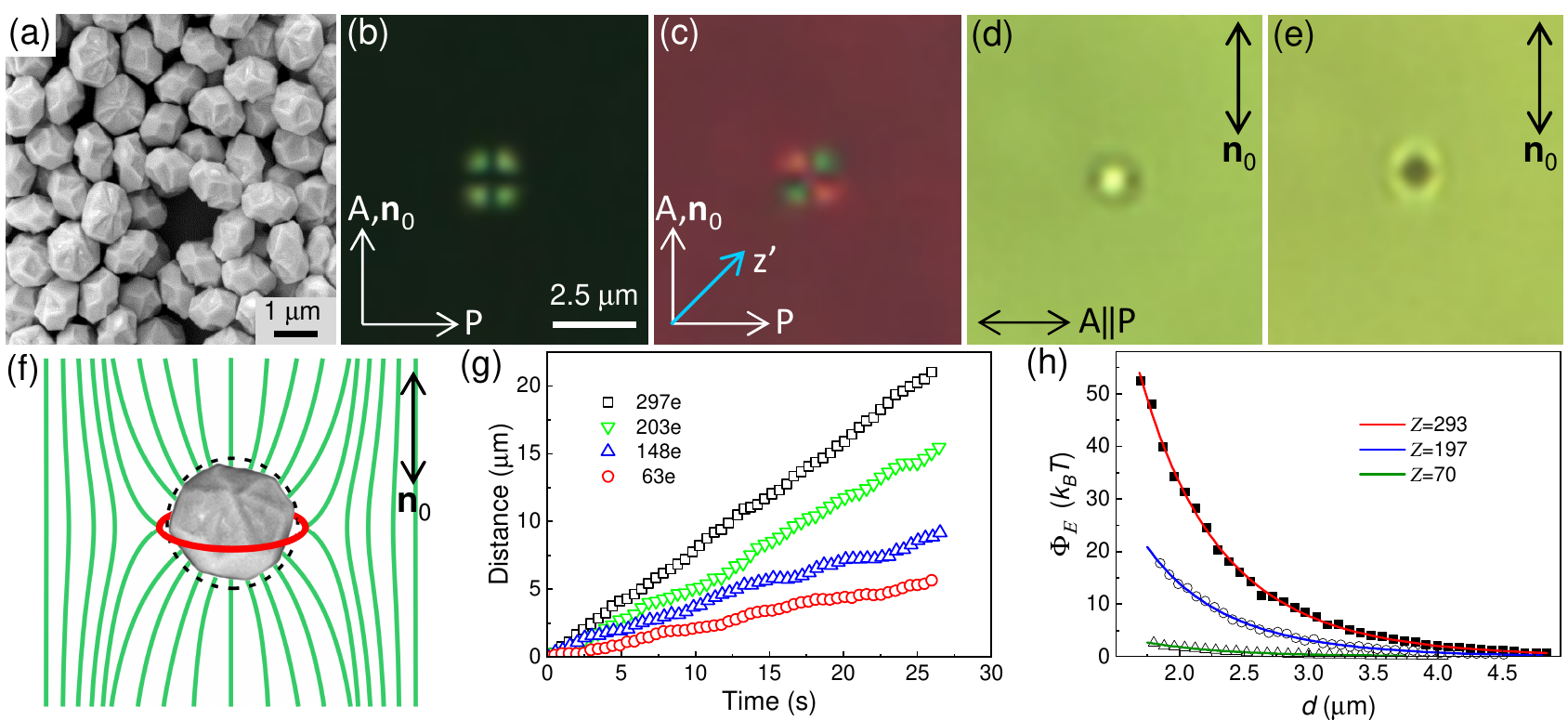}
\caption{{\bf Charged dumpling colloidal particles in nematic LC as model system for anisotropic charged colloids.} ({\bf a}) Scanning electron microscopy image of dumpling colloidal particles. ({\bf b-d}) Polarising and ({\bf e}) bright field microscopy textures of dumpling particles with homeotropic anchoring in a planar nematic cell between crossed ({\bf b, c}) and parallel ({\bf d}) polarizers P and A with ({\bf c}) and without ({\bf b}) a phase retardation plate $Z'$.  ({\bf f}) Schematic diagram of the director field {\bf n}({\bf r}) (green lines) around a dumpling particle treated for a homeotropic anchoring: a red circle around a particle indicates a singular defect loop ``Saturn ring" and $\mathbf{n}_0$ is a far-field director. ({\bf g}) Electrical effective charge of dumplings determined using electrophoresis. ({\bf h}) Repulsive electrostatic pair potential $\Phi_\mathrm{E}$ between dumpling colloids in an isotropic phase of a nematic LC with $\lambda_\mathrm{D}^\mathrm{I}\approx$ $925-959$ nm. }
\label{fig:expfig1}
\end{figure*}

In contrast, the host material anisotropy in {\color{black} screened electrostatic} interactions is a novel open challenge, centered at the question of how electrostatic screening is changed when the medium is characterised by a dielectric tensor, rather than a dielectric constant. Spatially dependent dielectric anisotropies occur sometimes in isotropic liquids, for example, near solid-water interfaces \cite{Netz:2011, Loche:2020}, but are usually confined in only a small region of space. A more general example of controllable anisotropic materials are liquid crystals (LCs), where the anisotropy is described by one dielectric coefficient along the primary  dielectric tensor axis, also called the director {\color{black} $\mathbf{n}\equiv-\mathbf{n}$}, and a second dielectric coefficient in the perpendicular direction. For a hypothetical, everywhere radial director around a colloidal sphere, only the dielectric tensor components projected in the radial direction contribute, and hence the Bjerrum length and consequently the Debye screening length are renormalised with a constant, and one can just {\color{black} use Eq. \eqref{eq:anisop} with $\mathcal{A}$ being constant}. However, a completely different situation arises when the director configuration around the particle does not have the same symmetry as the particle itself. The simplest non-trivial example would be a constant director field along the $z$ axis surrounding a spherical particle, and in this case the screening will become direction dependent, as we shall see here. Furthermore, as it is the more usual case in anisotropic nematic media with dispersed particles \cite{Poulin:1997,Musevic:2006,Lapointe:2009,Stebe:2013,Yuan:2015}, the director field is usually spatially dependent and varies in space because of various geometries, surface effects, and external fields, leading to rich and diverse elasticity-mediated anisotropic interparticle interactions. Multipole expansions have been used to describe elasticity-mediated colloidal interactions in  {\color{black} LCs}, drawing parallels to electrostatic interactions \cite{Lubensky:1998,Lev:2002,Yuan:2020}. In general, elasticity-mediated interactions in LCs are accompanied  by screened electrostatic and dispersive (London-van der Waals) interactions \cite{Mundoor:2016}; however, the previous studies of such colloidal systems with LC hosts were done for highly anisotropic rod- and disc-shaped particles, so that the role of the anisotropy of colloidal particles and that of the LC medium were not separated because of the particle's shape anisotropy and so far explored while probing phase behavior and self-assembly of colloidal superstructures \cite{Mundoor:2018,Mundoor:2019}. 


Here, we explore anisotropic colloidal interactions in electrostatically screened near-spherical charged colloids, to develop a generalised understanding of electrostatic interactions in colloids, subjected to and determined by the material dielectric anisotropy. Experimentally and theoretically, we use so-called charged dumpling particles (with almost spherical shape) as a charged colloidal model system, because they can become appreciably charged in a simple {\color{black} LC} such as 5CB (4-cyano-4'-pentylbiphenyl), with a weak-enough elastic interaction that allows competition with the electrostatic forces. We calculate the effective pair interaction under the assumption that elastic, dispersive and electrostatic interactions are additive, just like in the Derjaguin-Landau-Verwey-Overbeek (DLVO) theory of charged-screened spheres. Furthermore, the electrostatic part is treated within linear screening theory in combination with a far-field multipole expansion approach, on the same level that typically elastic colloidal interactions are treated. Last, we compare the theoretically calculated interactions with experiments, finding good qualitative agreement.

\section*{Results}
\subsection*{Charged colloidal dumpling particles dispersed in a nematic electrolyte as model system}
As our charged colloidal model system, we use particles (Fig. \ref{fig:expfig1}a) of ``dumpling-like" shape. Their rough shape and overall dimensions are close to those of a sphere with a diameter $2a=1 \ \mu m$. The hydrothermal synthesis and surface treatment of these particles allow us to control the colloidal charge in a rather broad range of values without issues (typical for other particles in nematic solvents) of nonuniformity of charging. These colloidal dumplings were dispersed in 5CB at low concentration ($<$ 1000 parts per million) to obtain well-separated colloidal particles. In Figs. \ref{fig:expfig1}b-e we show microscopy images of a single colloidal dumpling obtained in different imaging modes. The colloidal dumplings have homeotropic anchoring on their surfaces and the symmetry of resulting director  {\bf n}({\bf r}) distortions around particles (Fig. \ref{fig:expfig1}f) is of the “quadrupolar” type, with an encircling half-integer disclination loop (“Saturn ring”) \cite{Abbott:2000, Stark:2001}. The in-plane diffusion of the colloidal dumplings due to Brownian motion {\color{black} (fig. S8a,b and movie S1) } is anisotropic with respect to {\color{black} the LC far-field director } $\mathbf{n}_0$ with diffusion coefficients $D_{\parallel}/D_{\perp}$=1.49-1.54 {\color{black} (fig. S8c and movie S1)}, and this is close to theoretical predictions for spheres, $D_{\parallel}/D_{\perp}=1.72$ \cite{Stark:2001b}. We have prepared and used charged and uncharged particles in our experiments, where the effective charge was controlled in a broad range. The number of elementary charges {\color{black} $e$} on the particles surface $Z=0-350$ was determined using their electrophoretic motion between two in-plane electrodes placed perpendicular to $\mathbf{n}_0$ in a planar nematic cell (the uncharged particles did not move in response to applying an electric field). Dumpling charged particles were moving along $\mathbf{n}_0$ towards a negative electrode when a DC electric field was applied between the electrodes. The velocity of the particles depends on their charge {(Fig. \ref{fig:expfig1}g) } and the strength of the electric field. The displacement of particles was tracked using video microscopy, which allows us to estimate the effective charge $Ze$, from the balance of the Stokes viscous drag force and the electric force \cite{Mundoor:2016,Mundoor:2018,Mundoor:2019}. To probe only the electrostatic pair interactions between charged colloidal dumplings, we measured their pair interactions in the isotropic phase of 5CB, where the contribution due to LC elastic forces is eliminated. When brought nearby with the help of optical tweezers, colloidal dumplings repel from each other with a potential $\Phi_\mathrm{E}$ of tens of $k_\mathrm{B} T$. The effective charge number $Z$ and Debye screening length can also be extracted from experimental pair interactions (Fig. \ref{fig:expfig1}h) using the DLVO equation\cite{Derjaguin:1948, VerweyOverbeek}
\begin{equation}
\frac{\Phi_\mathrm{E}(d)}{k_\mathrm{B}T}=Z^2\lambda_\mathrm{B}\left[\frac{\exp\left(a/\lambda_\mathrm{D}^\mathrm{I}\right)}{1+a/\lambda_\mathrm{D}^\mathrm{I}}\right]^2\frac{\exp\left(-d/\lambda_\mathrm{D}^\mathrm{I}\right)}{d},
\label{eq:DLVO}
\end{equation}
with $k_\mathrm{B}T$ being the thermal energy, $\lambda_\mathrm{B}$ the Bjerrum length, $a$ the particle radius, $\lambda_\mathrm{D}^\mathrm{I}$ the (isotropic) Debye screening length and $d$ the center-to-center distance between particles. The effective charge numbers obtained by electrophoretic measurements (Fig. \ref{fig:expfig1}g) were in a good agreement with values obtained from the electrostatic interaction potential (Fig. \ref{fig:expfig1}h). The Debye screening lengths obtained from fitting the interaction potentials were within the range of $\lambda_\mathrm{D}^\mathrm{I}$=300-1000 nm measured for 5CB samples in our experiments using impedance spectroscopy.

Because of the effective elastic nature of the anisotropic LC host, also \emph{uncharged} colloidal particles interact via anisotropic elastic interactions \cite{Stark:2001}, which for our dumpling particles are of quadrupolar symmetry {\color{black} (fig. S9)} with elastic interaction potential $\Phi_\text{LC}(d,\theta)$ given by   
\begin{equation}
\Phi_\mathrm{LC}(d,\theta)=\frac{16}{9}\pi Kc^2\frac{9-90\cos^2\theta+105\cos^4\theta}{d^5}.
\label{eq:elastic}
\end{equation}
Note that the potential falls off as $\propto 1/{d^5}$ and depends on the angle $\theta$ between the uniform far-field $\mathbf{n}_0$ and $d$ which makes it strongly anisotropic, with the attraction direction at $\approx$ 40-50$^{\circ}$ {\color{black} (fig. S9a,b)}. We can extract the elastic pair potential {\color{black} (fig. S9d) } from {\color{black} the time dependent separation between two attracting particles  (fig. S9c)} and on the basis of the elasticity measurements (using the single elastic constant $K=8\cdot10^{-12}$ N), we find, for our system,  {\color{black} the elastic quadrupole moment} $c=0.1-0.2\ \mu$m$^3$.

\begin{figure*}[t]
\centering
\includegraphics[width=0.95\textwidth]{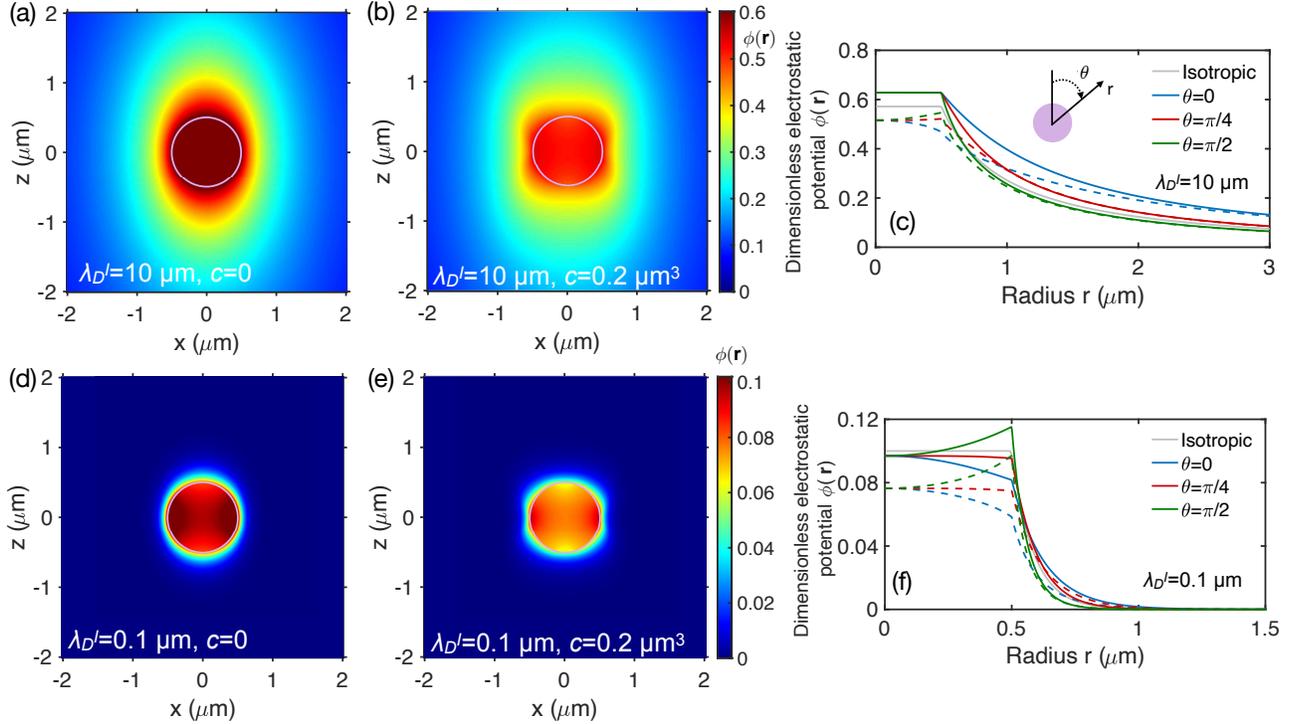}
\caption{{\bf Anisotropic electrostatic potential $k_\mathrm{B}T\phi({\bf r})/e$ for charged spherical colloidal particle in anisotropic dielectric host.}
({\bf a}) and ({\bf b}) Potential for large (isotropic) Debye screening length and ({\bf c}), ({\bf d}) short Debye screening length, with uniform director host $c=0$ (along the  $z$ direction) and with quadrupolar director material anisotropy  ($c\neq0$). ({\bf e}) and ({\bf f}) Electrostatic potential along selected directions from the particle center, at constant angle $\theta$ with respect to the $z$ axis. Full lines correspond to uniform, and dashed lines to quadrupolar material anisotropy $c=0.2\ \mu\mathrm{m}^3$. The gray line is the isotropic electrostatic potential line. For numerical parameters, we take particle radius $a=0.5\ \mu\mathrm{m}$ with constant charge $Z=50$ in a nematic electrolyte with dielectric properties $\epsilon_\perp=6$, $\epsilon_\parallel=19$ and $\bar{\epsilon}=10$.}
\label{fig:anisoelecpot}
\end{figure*}

\begin{figure*}[t]
\centering
\includegraphics[width=\textwidth]{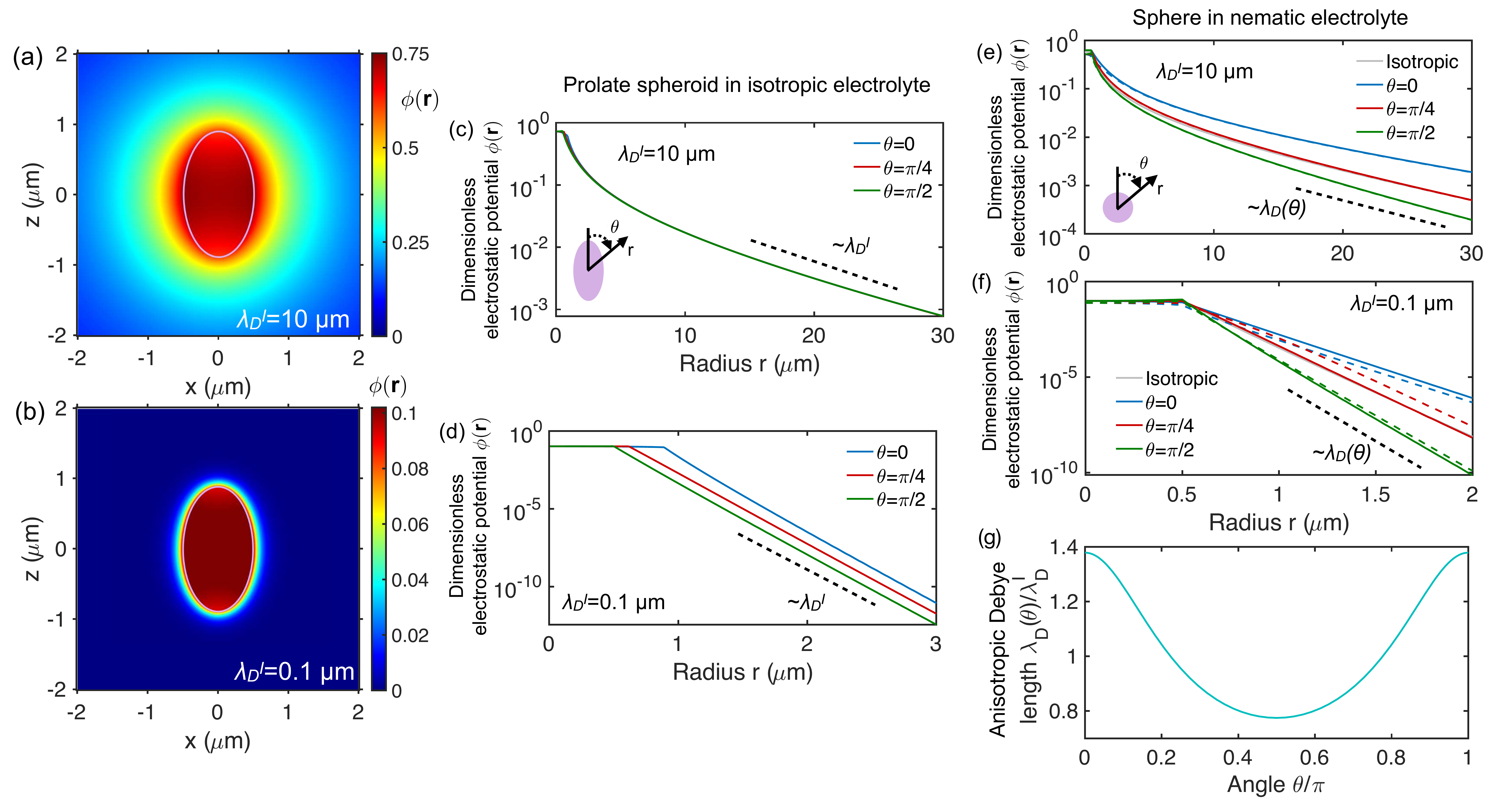}
\caption{{\bf Electrostatic potential of a charged prolate spheroid in an isotropic electrolyte compared with a charged sphere in a nematic electrolyte.}
({\bf a}) and ({\bf b}) Electrostatic potential for large and small Debye screening length, respectively around a prolate spheroidal particle in an isotropic electrolyte with minor radius $a=0.5$ $\mu$m and major radius $\sqrt{19/6}a$. In ({\bf c}) and ({\bf d}) we plot the potential along selected directions from the particle center at constant angle $\theta$ with respect to the $z$ axis in on log-linear scale. In ({\bf e}) and ({\bf f}) we replot Fig. \ref{fig:anisoelecpot}c, f on log-linear scale highlighting the angle-dependent screening length found in a nematic electrolyte as opposed to an isotropic electrolyte (({\bf c}) and ({\bf d})), described by a constant Debye length. ({\bf g}) Anisotropic Debye screening length $\lambda_\mathrm{D}(\theta)$ as calculated from Eq. \eqref{eq:anisoscreen} relative to the isotropic screening length $\lambda_\mathrm{D}^\mathrm{I}$ in different directions.}
\label{fig:prolate}
\end{figure*}

\subsection*{Electrostatic potential of single charged colloidal sphere in an anisotropic dielectric}

Electrostatic interactions between charged particles are conditioned by the profile of the electrostatic potential surrounding the particles. This quantity exhibits anisotropic screening as it is directly determined by the anisotropy of the host medium. We calculate this anisotropic electrostatic potential in the mean-field approach by using the Poisson-Boltzmann (PB) equation for the electrostatic potential in the nematic host with \emph{fixed} director field ${\bf n}({\bf r})$ surrounding the particle (see the Supplementary Materials),
\begin{equation}
\partial_i[\epsilon_{ij}({\bf r})\partial_j\phi({\bf r})]/\bar{\epsilon}=\kappa^2\sinh[\phi({\bf r})], \quad (r>a),
\label{eq:PB}
\end{equation}
where $k_\mathrm{B}T\phi({\bf r})/e$ is the electrostatic potential, $\bar{\epsilon}$ is the rotationally averaged dielectric constant of the nematic medium, and $\kappa^{-1}=\lambda_\mathrm{D}^\mathrm{I}$ is the isotropic Debye screening length used as a ``reference" decay length. Furthermore, we used the Einstein summation convention. Note that in the above, we assume that the dumpling particles can be approximated as spheres, have constant-charge boundary conditions with total homogeneously distributed charge $Ze$, and that {\color{black} $\mathbf{n}_0$} surrounding the particle has cylindrical symmetry. The host material anisotropy is given by the dielectric tensor as $
\epsilon_{ij}({\bf r})=\epsilon_\perp\delta_{ij}+\Delta\epsilon n_i({\bf r})n_j({\bf r}),$ with $\Delta\epsilon=\epsilon_\parallel-\epsilon_\perp$ being the dielectric anisotropy difference between dielectric tensor components projected parallel to the director $\epsilon_\parallel$ and perpendicular to the director $\epsilon_\perp$. We take, in accordance with our experiments, that ions cannot penetrate the particle with dielectric constant $\epsilon_\mathrm{p}=2$; hence, inside the particle, one has to solve the Laplace equation
\begin{equation}
\nabla^2\phi({\bf r})=0, \quad (r<a).
\label{eq:laplace}
\end{equation}
The anisotropic electrostatic potential [Eqs. \eqref{eq:PB} and \eqref{eq:laplace}] is numerically calculated using COMSOL Multiphysics software exploiting the cylindrical symmetry, as shown in Fig. \ref{fig:anisoelecpot}. We provide results for two material anisotropic regimes, one with uniform director field ${\bf n}={\bf e}_z$ and second -the realistic one for our experiments- a director field with elastic quadrupole distortions  ${\bf n}=[{\bf e}_z+(2cz\rho/r^5){\bf e}_\rho]/|{\bf e}_z+(2cz\rho/r^5){\bf e}_\rho|$ (as shown in Fig.~\ref{fig:expfig1}f). The quadrupole is derived from a multipolar expansion for the Saturn-ring configuration \cite{Stark:2001}.

For the uniform director field  (Fig. \ref{fig:anisoelecpot}a) and a Debye screening length larger than the particle size, the diffuse screening cloud has a prolate-spheroidal like shape with the long axis coinciding with the $z$ axis, whereas upon considering the full quadrupolar distortion,  we see that the electric potential profile (i.e. the double layer) gets  distorted close to the region of the Saturn ring defect (Fig. \ref{fig:anisoelecpot}b), but again evolves to the prolate spheroid shape further away from the particle. {\color{black} Furthermore, it turns out that the spheroidal double layer has an aspect ratio for the major to minor axis equal to $\sqrt{\epsilon_\parallel/\epsilon_\perp}$.} Fig. \ref{fig:anisoelecpot}c shows the electrostatic potential along selected directions from the particle at constant angle $\theta$ with respect to the $z$ axis, which can be compared to the isotropic electrostatic potential, showing a similar magnitude. In Figs. \ref{fig:anisoelecpot}d-f, the electrostatic potential for the regime of Debye screening length smaller than the particle size is shown for a uniform director (Fig. \ref{fig:anisoelecpot}d) and quadrupolar director (Fig. \ref{fig:anisoelecpot}e); the electrostatic potential along constant $\theta$ is shown in Fig. \ref{fig:anisoelecpot}f. At these short Debye lengths, the electrostatic potential inside the particle becomes strongly inhomogeneous, and although the ions are closer to the particle, the electrostatic potential is still strongly anisotropic. 

{\color{black}
Figure \ref{fig:prolate} shows that the electrostatic potential of a spherical particle with a prolate spheroidal double layer in a nematic electrolyte behaves differently from a charged spheroidal particle with aspect ratio $\sqrt{\epsilon_\parallel/\epsilon_\perp}$ in an isotropic electrolyte. Whereas the diffuse screen cloud seems to follow the shape of the spheroidal particle sufficiently close to the particle (Fig. \ref{fig:prolate}a,b), a graph on log-linear scale along a selection of cuts through the particle reveals that only the ``amplitude" is anisotropic for sufficiently small $\lambda_\mathrm{D}^\mathrm{I}$. However, for all bulk ion concentrations, the screening length is independent of the angle. We already alluded this well-known result in Eq. \eqref{eq:anisop}. In contrast, the sphere in a nematic shows --with and without-- topological defects that there is an angle-dependent screening length (Figs. \ref{fig:prolate}e,f). Such an anisotropy occurs whenever the director on the particle surface (in this example spherical) has a different symmetry than the far-field director field (cylindrical).}

\subsection*{Analytical expressions for the electrostatic potential}
{\color{black}To determine the  anisotropic electrostatic potential, we first derive an analytical expression by mapping the charged particle to an ion-penetrable spherical shell with radius $R$ and surface charge density $e\sigma_\mathrm{S}$, followed by a multipole expansion. The approach is more extensively explained in the Materials and Methods, the Supplementary Materials and Ref. \cite{Everts:2020b}.}
The calculated electric potential is 
\begin{align}
\phi({\bf r})=\frac{\alpha\gamma^2Z\lambda_\mathrm{B}^\mathrm{I}}{\sqrt{\epsilon_\perp^2\epsilon_\parallel/\bar{\epsilon}^3}}\Big[G_\mathrm{m}({\bf r})+&\frac{(\gamma a)^2}{6}G_\mathrm{q}({\bf r})\nonumber \\
&+\frac{(\gamma a)^4}{120}G_\mathrm{h}({\bf r})+...\Big],
\label{eq:epot}
\end{align}
as expressed with multipolar basis functions $G_i({\bf r})$ $(i=\mathrm{m,q,h},..)$ (see the Supplementary Materials for their expressions). The higher-order multipoles become more important at higher salt concentrations (smaller $\lambda_\mathrm{D}^\mathrm{I}$), and the parameters $\alpha$ and $\gamma$ also depend on $a/\lambda_\mathrm{D}^\mathrm{I}$. The salt-dependent parameters $\alpha=\sigma_\mathrm{S}/\sigma>1$ and $\gamma=R/a<1$ can be determined using a fit to the numerically obtained surface potential. By fitting only the surface potential, it is found that the integral expression thatis obtained before performing the multipole expansion, is practically \emph{numerically exact} for $r>a$ in a wide range of salt concentrations, see the Supplementary Materials.

As is usual for multipole expansions, Eq. \eqref{eq:epot} fails at short distances, but at large enough $r=|{\bf r}|$ it captures the proper angle dependence, given that enough multipoles are taken into account (see the Supplementary Materials). For example, up until hexadecapolar order, Eq. \eqref{eq:epot} is numerically exact up until $a/\lambda_\mathrm{D}^\mathrm{I}\sim2$ ($< 5\%$ deviation), while at higher salt concentrations truncation at the hexadecapolar order turns out to be not sufficient. As an example, for $a/\lambda_\mathrm{D}^\mathrm{I}\sim5$, we see \emph{even} asymptotically far from the particle, that the deviation is $10-60\%$, depending on the angle with the director, see the Supplementary Materials. Last, $Z$ is the actual charge number for $|\phi({\bf r})|\ll1$, but for high electrostatic potentials, $Z$ should be interpreted as a renormalised charge density, similar to what is known in ``isotropic" charged colloids \cite{Alexander:1984, Trizac:2003}.

{\color{black} From Eq. \eqref{eq:epot} we derive the asymptotic scaling
\begin{equation}
\phi({\bf r})\sim Z\mathcal{A(\theta)}\frac{\exp\left[-r/\lambda_\mathrm{D}(\theta)\right]}{r/\lambda_\mathrm{B}(\theta)}, \quad (r\rightarrow\infty),
\label{eq:monopole}
\end{equation}
with the anisotropy function for uniform director fields
\begin{align}
\mathcal{A}(\theta)=\alpha\gamma^2\Bigg\{&1+\frac{(\gamma\kappa a)^2}{6}\bar{\epsilon}^3\left(\frac{\cos^2\theta}{\epsilon_\parallel}+\frac{\sin^2\theta}{\epsilon_\perp}\right) \nonumber\\
&\times \left(\frac{\cos^2\theta}{\epsilon_\parallel^2}+\frac{\sin^2\theta}{\epsilon_\perp^2}\right)+\mathcal{O}\left[(\kappa a)^4\right]\Bigg\},
\end{align}
}
angle-dependent Bjerrum length,
\begin{equation}
\frac{\lambda_\mathrm{B}(\theta)}{\lambda_\mathrm{B}^\mathrm{I}}=\frac{\bar{\epsilon}}{\sqrt{\epsilon_\perp(\epsilon_\parallel-\Delta\epsilon\cos^2\theta)}},
\label{eq:B}
\end{equation}
and angle-dependent Debye screening length
\begin{equation}
\frac{\lambda_\mathrm{D}(\theta)}{\lambda_\mathrm{D}^\mathrm{I}}=\sqrt{\frac{\epsilon_\perp\epsilon_\parallel}{\bar{\epsilon}(\epsilon_\parallel-\Delta\epsilon\cos^2\theta)}},
\label{eq:anisoscreen}
\end{equation}
with for $\Delta\epsilon>0$ ($\Delta\epsilon<0$) a maximum (minimum) at $\sqrt{\epsilon_\parallel/\bar{\epsilon}}$ and a minimum (maximum) at $\sqrt{\epsilon_\perp/\bar{\epsilon}}$. Note that $\lambda_\mathrm{D}(\theta)/\lambda_\mathrm{D}^\mathrm{I}\neq\sqrt{\lambda_\mathrm{B}^\mathrm{I}/\lambda_\mathrm{B}(\theta)}$, as one would maybe naively think based on $\lambda_\mathrm{D}^\mathrm{I}=(8\pi\lambda_\mathrm{B}^\mathrm{I}\rho_s)^{-1/2}$.  We plotted Eq. \eqref{eq:anisoscreen} in Fig. \ref{fig:anisoelecpot}g for $\Delta\epsilon>0$, and from this the shape of the double layer can be understood. Had we taken $\Delta\epsilon<0$, the double layer would have had an oblate-spheroidal shape.

{\color{black}
Eq. \eqref{eq:monopole} reveals why electrostatic screening in an anisotropic medium behaves markedly different from screening of anisotropic particles in an isotropic medium. For $r\rightarrow\infty$, the particle anisotropy can be fully accounted for with the anisotropy function $\mathcal{A}(\theta)$, see Eq. \eqref{eq:anisop}, whereas anisotropic media are described not only by an anisotropy function (even for spheres with non-vanishing volume), but also by an angle-dependent Debye and Bjerrum length, see Eqs. \eqref{eq:B} and \eqref{eq:anisoscreen}, respectively. The anisotropic screening length is highlighted in Figs. \ref{fig:prolate}e, f, and is the relevant decay length even in the presence of defects. Furthermore, Figs. \ref{fig:prolate}e, f show in accordance with Eq. \eqref{eq:monopole} that the anisotropy function is only relevant for a sufficiently large $\kappa a$, giving rise to larger amplitude differences of the electrostatic potential for varying $\theta$ and at fixed $r$. Furthermore, Fig. \ref{fig:prolate}f suggests that defects alter the form of $\mathcal{A}(\theta)$, but not that of $\lambda_\mathrm{D}(\theta)$, which is determined only by the symmetry of the far-field director. Therefore, we hypothesise that defects asymptotically behave as if they change the \emph{particle} anisotropy but not the \emph{medium} anisotropy.}

Last, while assuming that the strength and nature of surface boundary conditions on particle surfaces do not change with adding ions, we evaluate the salt-dependent renormalisation parameters $\alpha$ and $\gamma$ for some values of $a/\lambda_\mathrm{D}^\mathrm{I}$ (more information in the Supplementary Materials). For $a/\lambda_\mathrm{D}^\mathrm{I}=0.5$, we find $\gamma=1$ and $\alpha=1.06$; for $a/\lambda_\mathrm{D}^\mathrm{I}=1$, we find $\gamma=0.99$ and $\alpha=1.2$; and for $a/\lambda_\mathrm{D}^\mathrm{I}=2$, we find $\gamma=0.988$ and $\alpha=1.43$.  { \color{black} With these quantities, we can use Eq. \eqref{eq:epot} in the relevant experimental parameter regime, and the same parameters enter the expression for the effective far-field pair potential (see below).}

\begin{figure*}[t]
\centering
\includegraphics[width=0.95\textwidth]{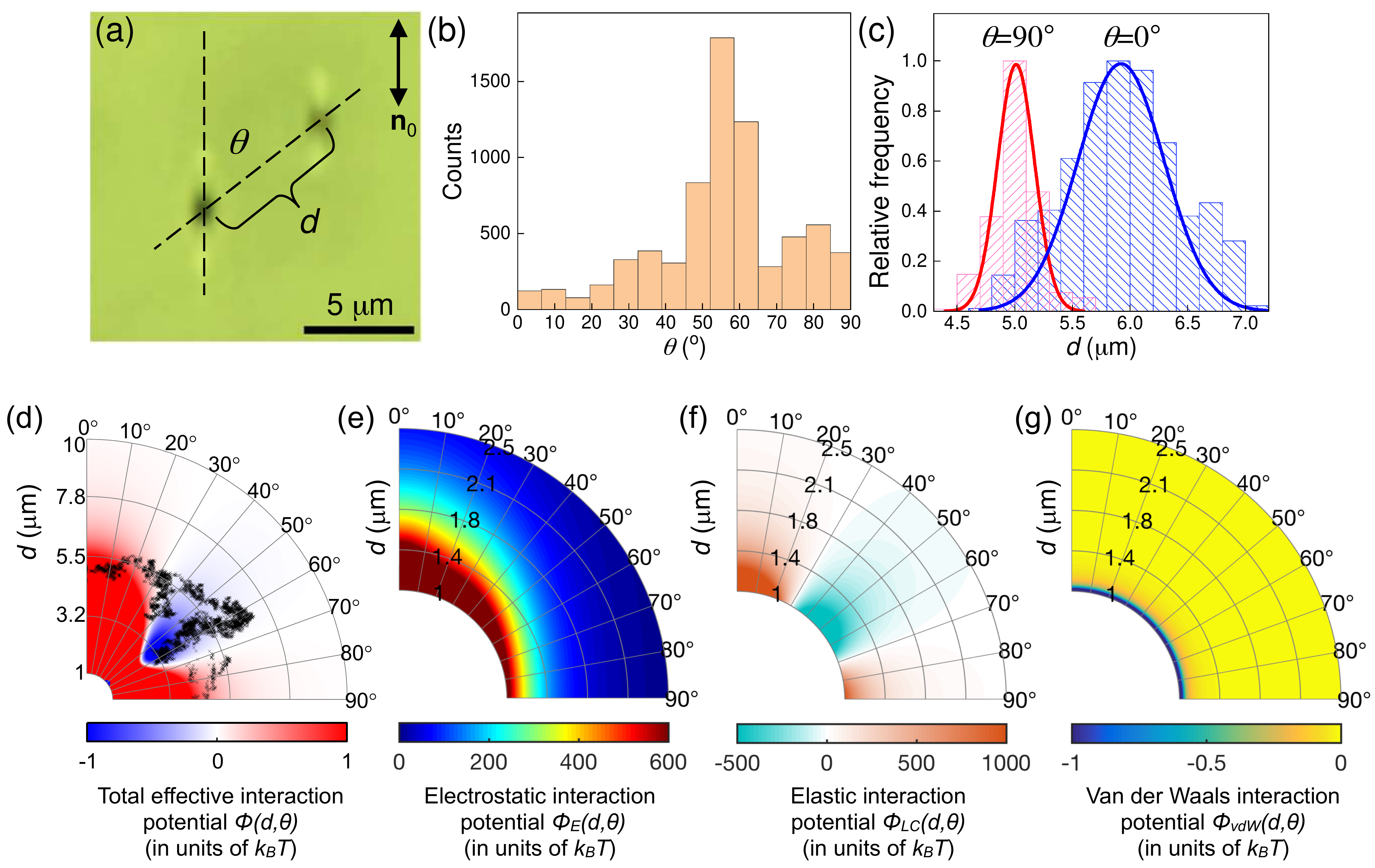}
\caption{{\bf Pair interactions of charged dumpling particles in a nematic LC.} ({\bf a}) Bright-field micrograph of two interacting colloidal dumplings. { \color{black} ({\bf b}) Histogram showing the preferred angle $\theta$ between the center-to-center separation vector and the far-field director for a pair of two highly charged particles. ({\bf c}) Histograms showing preferred separations between two highly charged particles ($\sim$300$e$ per particle) when their separation vector is at $\theta\approx0^\circ$ (blue) and $\theta\approx90^\circ$ (red). } ({\bf d}) Analytically calculated total effective potential as the sum of contributions from: ({\bf e}) anisotropic electrostatic interactions, ({\bf f}) anisotropic elastic interactions and ({\bf g}) van der Waals dispersion interactions. The interaction potentials are calculated for particle radius $a=0.5\ \mu\mathrm{m}$, particle charge number $Z=750$, elastic quadrupole moment $c=0.1\ \mu\mathrm{m}^3$, Hamaker constant $A_\mathrm{H}=1\ k_\mathrm{B}T$, and nematic electrolyte with isotropic Debye screening length $\lambda_\mathrm{D}^\mathrm{I}=0.5\ \mu\mathrm{m}$. {\color{black} Black crosses in {\bf d} show the overlaid angular dependent separations between two interacting charged particles: particles do not come in direct contact at any angle. Note that experimental preferred angles and separations are consistent with the calculated competing electrostatic and elastic potentials, with the minimum of the total effective interaction potential at $\sim55^\circ$.} }
\label{fig:expfig3}
\end{figure*}

\subsection*{Anisotropic  electrostatic and elastic pair interactions}
Charged colloidal dumplings interact in the nematic LC {\color{black} (Fig. \ref{fig:expfig3}a and movie S2)}, both via anisotropic electrostatic and elastic interactions, each with a different anisotropic profile. Generally, the electrostatic interaction is repulsive, whereas the nematic elastic interaction has regions (directions) of attraction and regions of repulsion. If the charge of the particles is high enough {\color{black} ($\sim300e$)}, then the electrostatic repulsion can counterbalance the elastic attraction at any angle,  and the two colloidal particles {\color{black} stay separated from each other without getting into full contact (Fig. \ref{fig:expfig3}) at any $\theta$. The orientation of the separation vector between particles is primarily $\approx$55$^{\circ}$ (Fig. \ref{fig:expfig3}b)} relative to $\mathbf{n}_0$. {\color{black} We note that for this histogram, we mapped the angles to the first quadrant because of the quadrupolar symmetry of the system.} {\color{black} Histograms of the center-to-center distance between particles (Fig. \ref{fig:expfig3}c) show that, for selected angles $\theta\approx $0$^{\circ}$ and 90$^{\circ}$ where we used laser tweezers to release the particle pair along these angles, the separation is $d\approx6$ and $5 \ \mu$m, respectively, which is consistent with the anisotropy of the calculated electrostatic repulsion force and overall colloidal interactions in this system (Fig. \ref{fig:expfig3}e)}. This difference in the steady-state $d$ depends on the position of the two particles relative to {\color{black} $\mathbf{n}_0$}, which can result from the anisotropy in the charge distribution around colloidal dumplings (Fig. \ref{fig:anisoelecpot}). Histograms of the angle $\theta$ and an overlaid polar plot of separations between two freely moving particles (Figs. \ref{fig:expfig3}b,d) show that there {\color{black} is a preferred orientation for a separation $d$ at $\approx$55$^{\circ}$} (again mapping all the data to the first quadrant), which indicate an equilibrium distance and orientation of a particle pair resulting from a competition of anisotropic elastic attraction and anisotropic electrostatic repulsion (Fig. \ref{fig:expfig3}d-g), as we shall explore theoretically below. On the other hand, if the electric charge at the dumpling surface is sufficiently small  {\color{black} ($\sim150e$)}, then the elastic attraction is dominant and two colloidal particles attract and get to the full surface-to-surface contact as elastic quadrupoles {\color{black} (fig. S10), similar to colloidal dumplings without charge {\color{black} (fig. S9 and movie S3)}}.


The effective pair interaction between anisotropic charged colloids is determined theoretically, by splitting --- in the spirit of the DLVO theory --- the total interaction potential $\Phi(d,\theta)$, as the sum of screened electrostatic $\Phi_\mathrm{E}(d,\theta)$, van der Waals $\Phi_\mathrm{vdW}(d)$, and, because we are in a nematic host, the nematic elastic interactions $\Phi_\mathrm{LC}(d,\theta)$
\begin{equation}
\Phi(d,\theta)=\Phi_\mathrm{E}(d,\theta)+\Phi_\mathrm{vdW}(d)+\Phi_\mathrm{LC}(d,\theta).
\label{eq:totalpoten}
\end{equation}
The van der Waals interaction can be derived using Hamaker-de Boer theory
\begin{equation}
\Phi_\mathrm{vdW}(d)=-\frac{A_\mathrm{H}}{3}\left[\frac{a^2}{d^2-4a^2}+\frac{a^2}{d^2}+\frac{1}{2}\ln\left(1-\frac{4a^2}{d^2}\right)\right],
\end{equation}
which fails at center-to-center distances close to contact ($d\approx 2a$) where the (quantum-mechanical) Born repulsion becomes important, and for large $d$ where relativistic effects become important. For this interaction, the anisotropy enters only the Hamaker constant, but not in the coordinate-dependent part of the expression, assuming the nonrelativistic limit \cite{Sarlah:2001}. Last, the elastic interaction $\Phi_\mathrm{LC}(d,\theta)$ is given by the quadrupolar far-field expression Eq. \eqref{eq:elastic}.
\begin{figure*}[t]
\centering
\includegraphics[width=0.95\textwidth]{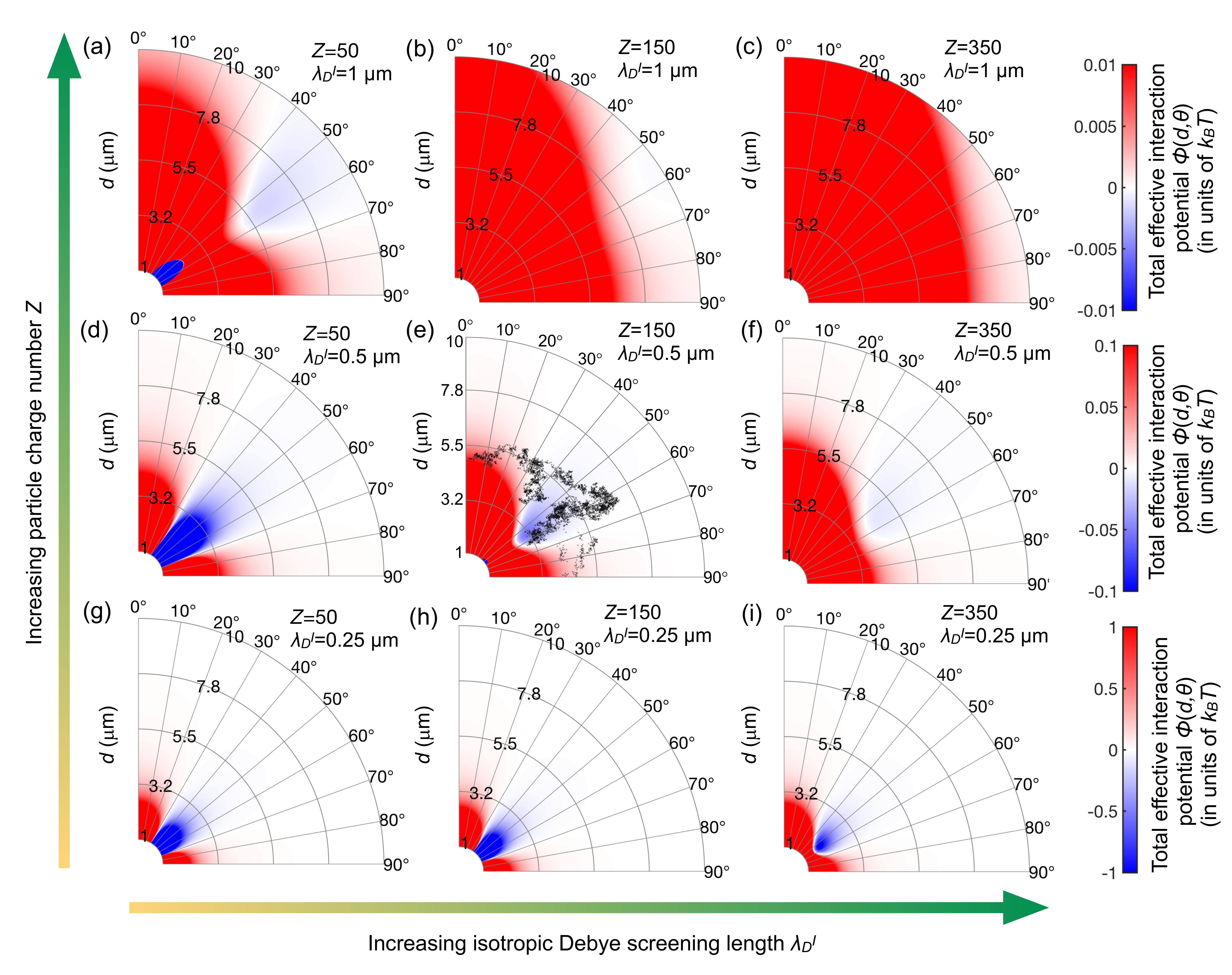}
\caption{{\bf Effect of particle charge and salt concentration on total interaction pair-potential of spherical charged colloidal particles with quadrupolar nematic dielectric anisotropy.} ({\bf a-i}) The total potential is calculated for particle radius $a=0.5 \ \mu\mathrm{m}$ and elastic quadrupole magnitude $c=0.02\ \mu\mathrm{m}^3$ for varying particle charge $Z$ and different screening lengths $\lambda_\mathrm{D}^\mathrm{I}$. Note the difference in colour scales for different screening lengths. {\color{black} Black dots in {\bf e} show experimental results, i.e., relative particle separations, overlying the theoretically calculated interaction potential in the relevant material parameter regime ($Z=150$, $\lambda_\mathrm{D}^\mathrm{I}=0.5\ \mu m$). Note  that particles do not come in direct contact at any angle.}}
\label{fig:totalintvarpar}
\end{figure*}

We determine the effective anisotropic screened electrostatic interaction $\Phi_\mathrm{E}(d,\theta)$ from the asymptotic expression of the electrostatic potential [Eq. \eqref{eq:epot}] within the linear superposition approximation (LSA, for full derivation, see the Supplementary Materials). The anisotropic screened electrostatic interaction reads 
\begin{align}
\frac{\Phi_\mathrm{E}(d,\theta)}{k_\mathrm{B}T}=\frac{\alpha^2\gamma^4Z^2\lambda_\mathrm{B}^\mathrm{I}}{\sqrt{\epsilon_\perp^2\epsilon_\parallel/\bar{\epsilon}^3}}\Big[&G_\mathrm{m}(d,\theta)+\frac{(\gamma a)^2}{3}G_\mathrm{q}(d,\theta)+ \nonumber \\
&\frac{2}{45}(\gamma a)^4 G_\mathrm{h}(d,\theta)+...\Big],
\label{eq:supadupa}
\end{align}
now treated on the same level of (multipolar) approximation as the elastic interactions, where $G_\mathrm{m,q,h}(d,\theta$), $\alpha$ and $\gamma$ are the same as the ones derived for the single-particle case, see Eq. \eqref{eq:epot}. Note also, that in the limit of low salt concentration, only the monopole term is relevant and the above equation reduces to the anisotropic Yukawa form
\begin{equation}
\frac{\Phi_\text{E}(d,\theta)}{k_\mathrm{B}T}=\alpha^2\gamma^4Z^2\lambda_\mathrm{B}(\theta)\frac{\exp[-d/\lambda_\mathrm{D}(\theta)]}{d},
\end{equation}
which is reminiscent of the well-known standard isotropic DLVO potential Eq. \eqref{eq:DLVO}.

In Fig. \ref{fig:expfig3}d we show the total analytically calculated interaction potential for parameters, in qualitative range of our experiments. The position of a local minimum of several $k_\mathrm{B}T$s is found at  $\theta \sim 55^\circ$, which is in good qualitative agreement with experiments (Fig. \ref{fig:expfig3}b). {\color{black} The position of the theoretically calculated local minimum is also in good agreement with the overlaid experimentally determined orientations of the separation vector (Fig. \ref{fig:expfig3}d) where a pair of interacting particles reside primarily also at $\approx 55^\circ$.} In Fig. \ref{fig:expfig3}e, we show the contributions of the anisotropic electrostatic effective interaction, {\color{black} which also shows a good agreement with the experiment (Fig. \ref{fig:expfig3}c)}; in Fig. \ref{fig:expfig3}f, the elastic interaction; and in Fig. \ref{fig:expfig3}g the van der Waals interaction. The van der Waals interaction is of shortest range and does not contribute substantially to the total interaction potential, whereas elastic and screened electrostatic potentials clearly compete, and it is their detailed balance that determines the overall inter-particle potential. Last, note that if we had used the isotropic DLVO potential [Eq. \eqref{eq:DLVO}] to calculate the total potential (and using same material parameters), the predicted local minimum would shift to an angle $\theta\sim 49^\circ$, which underlines the clear role of the electrostatic anisotropy. 

The total interaction potential Eq. \eqref{eq:totalpoten} exhibits a range of possible qualitatively different inter-particle interaction regimes, as we show in Fig. \ref{fig:totalintvarpar}, which depend on the particle charge $Z$ and the host electrolyte screening length $\lambda_\mathrm{D}^\mathrm{I}$ (salt concentration). Notably, we calculate the total regimes for a smaller elastic interaction $c=0.02 \ \mu\mathrm{m}^3$ than in our experiments ($c\sim 0.2 \ \mu\mathrm{m}^3$), which makes the anisotropic electrostatic DLVO-type interaction of more similar magnitude as the nematic elasticity at the reported particle charges, which are lower than the one chosen in Fig. \ref{fig:expfig3}d. We want to stress, however, that the electrostatics is based on a far-field multipole expansion in combination with the LSA; hence, our theory underestimates the repulsion when the double layers of the particles overlap at sufficiently low salinity and low particle separations {\color{black} (fig. S5)}, as is also common in isotropic DLVO theory \cite{Carnie:1993}, not to mention because of the currently unknown nonadditive effects between elasticity and electrostatics, or close-approach elastic effects. 

In Figs. \ref{fig:totalintvarpar}a-c we show the situation when the screening length is larger than the particle size. For low enough charges, a global minimum located at approximately $49^\circ$ is separated from a shallow local minimum with depth $\sim0.001 \ k_\mathrm{B}T$. Both local and global minimum disappear when the particle charge is increased (Figs. \ref{fig:totalintvarpar}b and c), and the interaction potential is repulsive for all distances and directions. 
For Debye lengths similar to the particle size, we see that higher charges are needed to overcome the attractive elastic interaction. At the same particle charge but smaller screening length (Fig. \ref{fig:totalintvarpar}d), the electrostatic interaction is too weak to overcome the elastic interaction, resulting in a purely attractive direction in the interaction potential. Such a situation is reminiscent of what we also observed experimentally (but for slightly different exact parameters) in {\color{black}fig. S10}, where a low charge results in particle coagulation. When increasing $Z$ further, again a local minimum (Fig. \ref{fig:totalintvarpar}e) that is deeper than the low-screening case emerges. {\color{black} This local minimum is in a good agreement with experimentally determined angular dependent separations between interacting particles overlaid over the calculated data, even though we choose a smaller value for the elastic quadrupole than was experimentally measured (fig. S9); therefore, a lower particle charge is needed to generate the metastable local minimum.} The depth of this minimum becomes smaller when $Z$ increases even further (Fig. \ref{fig:totalintvarpar}f). The trend that deeper attractive minima can be attained for larger salt concentrations, but that higher charges are needed, is something that we also see for double layers smaller than the particle size
(Figs. \ref{fig:totalintvarpar}g-i). When comparing Fig. \ref{fig:totalintvarpar} with Fig. \ref{fig:expfig3}, we see that increasing the strength of elastic interaction at the expense of higher $Z$ also gives rise to deeper attractive minima.

\section*{Discussion}
Summarising, we introduced a screened electrostatic colloidal model system that can become appreciably charged in a nematic LC, with particles of almost spherical shape. Theoretically, we discussed that dielectric anisotropy of the colloidal host, given in nematic fluids by the director field, gives rise to anisotropic screening of the electrostatic potential when there is a mismatch between the nematic and particle symmetry, and in turn, also to the electrostatic effective pair interaction. In our system of nematic colloids, the screened electrostatic interaction is inherently combined with effective nematic elastic interactions, which leads to different interaction regimes where particles are (i) repelled for all distances and angles, (ii) are subjected to a weak local minimum of $\lesssim 1 k_\mathrm{B}T$ such that they can still move because of thermal fluctuations {\color{black} (Figs. \ref{fig:expfig3}b,d)} and (iii) are dominated by the elastic interactions with distinct attractive and  repulsive directions {\color{black} (fig. S10)}. In our experiments with charged colloidal dumplings, we have observed regime (ii) at {\color{black}$Ze\sim300e$} and regime (iii) at {\color{black}$Ze\sim150e$}, whereas to access for regime (i) we would need to achieve even higher particle charges and/or more deionised samples. However, we note that the regime (i) was achieved in experiments with highly shape-anisotropic particles in Ref. [41], where the small diameter of rod-like colloidal particles allowed for fully overpowering elastic interactions by the electrostatic ones. Our model is consistent with the notion that controlling particle dimensions provides the means of shifting the balance between the elastic and electrostatic forces at given colloidal surface charge and ionic conditions.

We have shown, within an experimentally accessible parameter range, that even effectively spherical particles exhibit strongly anisotropic electrostatic interactions in LCs. Experimental results indicate that both elastic and electrostatic interactions of charged particles in nematic LCs are relatively long ranged and anisotropic with respect to the director, so that the colloidal behaviour depends on their interplay. While charging could be controlled from neutral to hundreds of elementary effective charges per single particle, we could show that the inter-particle forces could be dominated by elasticity or by electrostatics in the two limiting regimes, with both elastic and electrostatic interactions being highly anisotropic. While we focused on thermotropic nematics with accessible range of host medium's Debye screening lengths in the range of 300 to 1000 nm and on colloidal particles with the accessible range of surface charges (0 to 350)$e$, our findings do indicate a plethora of colloidal behaviours arising from the interplay of electrostatic and elastic interactions with salient anisotropic features, consistent with theoretical modeling. 

This study can be, in the future, extended further by exploiting the elasticity and electrostatics interplay for lyotropic water-based LC colloidal systems, where Debye screening lengths can be much shorter, on the order of several nanometers, as well as highly deionized LCs that potentially could allow for accessing the range of Debye screening lengths from several nanometers to ~10 $\mu$m. As shown in Fig. \ref{fig:totalintvarpar}, tuning the screening length can change the relative position of a local minimum in the effective pair potential. Moreover, we note that lyotropic systems, might have additional features that our theory does not explore yet, being often a five-component mixture (an isotropic solvent such as water, LC particles, cations, anions, and the larger colloidal particles), compared with the thermotropic four-component suspension in this paper (LC, cations, anions and colloidal particles). Note that lyotropic systems can also have a dielectric anisotropy that couples the director with electrostatics \cite{Sonin:1987}, just as in the thermotropic systems under consideration here. Furthermore, when the liquid-crystal lyotropic particles are smaller colloidal (nano)particles, we envisage \emph{tuning} of the dielectric, elastic,  and possibly flexoelectric properties of the nematic host medium by changing the particle functionality.  Tuning of elastic properties by charged nanoparticles as a function of particle charge and salt concentration has already been explored theoretically \cite{Drwenski:2016}. Moreover, lyotropic systems can be made active, giving rise to possibly new unexplored hydrodynamic-electrokinetic active processes, which may be interesting also in a biological setting.

From the point of view of particle designs, these studies could be extended to patchy particles with different densities or even signs of charges, potentially allowing for different electrostatic multipoles, whereas our study showed that homogeneously charged spheres only give rise to even anisotropic Yukawa multipoles. Furthermore, while highly anisotropic disc- and rod-shaped particles have already been studied \cite{Mundoor:2016,Mundoor:2018, Mundoor:2019}, there is a considerable range of possibilities in defining colloidal behaviour also by the particle shape and surface treatment for different boundary conditions for the LC director at colloidal surfaces.

As further main theoretical results, we derived asymptotic theoretical expressions for the electrostatic potential and the resulting pair interaction for homogeneous director configurations, which highlights an anisotropy not only in the screening length but also in the prefactor, where the latter also occurs for anisotropic particles in isotropic solvents. Both anisotropies together predict that local minima in the total pair potential are shifted {\color{black} (Figs. \ref{fig:expfig3}b,d}) in terms of the equilibrium angle compared with an isotropic electrostatic interaction, in line with our experimental observations. Last, one obvious extension to the theory is to numerically investigate the effect of close distances between particles and relaxing the requirement of additivity in the pair potential by coupling the full Landau-de Gennes theory in terms of the tensor order parameter with electrostatics. We will explore this in future work.

More generally, our work contributes to the generalisation and extension of the DLVO interactions to  ubiquitous anisotropic soft matter systems, such as complex fluids and anisotropic colloids. While nematic colloids formed by a thermotropic LC host and near-spherical colloidal inclusions provide validation of our theoretical findings, the concepts introduced here can be applied in biological contexts of highly structured biological cell interior and membranes, active matter systems with the additional forms of anisotropy stemming from activity, lyotropic LCs with varying degrees of orientational and partial positional ordering, ionic fluids, and so on. While the experiments and model focused on even anisotropic Yukawa multipoles formed by like-charged spherical particles, future studies can extend our concepts to odd anisotropic Yukawa multipoles via a heterogeneous surface charge distribution or non-spherical particle shape on the electrostatic side of the spectrum, and to elastic monopoles through hexadecapoles and higher-order multipoles on the elastic side. It will be of interest to consider further the {\color{black}(non-additive)} effects of various topological defects on counterion distributions, the role of flexoelectricity and surface polarisation, surface anchoring effects, {\color{black} charge regulation}, flow \cite{Takeuchi:2010}, and how similar concepts work in LC mesophases with different point group symmetries and partial positional ordering. {\color{black} Specifically, one can expect to have the double layer driven realignment of the nematic and even electrostatically controlled surface anchoring transitions as function of salt concentration and interparticle distance in the regime where the dielectric torque is of similar magnitude as the elastic torque \cite{Shah:2001, Onuki:2009, Mundoor:2019, Everts:2020}. Furthermore, flexoelectric topological defects can substantially alter charge distributions on the colloidal particles \cite{Ravnik:2020}. These are just a few of the many examples of additional higher-order phenomena of which the effects on the effective interparticle potential are left for future studies.} Overall, our findings will contribute to the soft matter toolkit for forming colloidal composite materials with pre-engineered structure and composition of the constituent building blocks, {\color{black} and will also help further in understanding non-equilibrium phenomena in these systems \cite{Lavrentovich:2017}.}

\section*{Materials and methods}
\subsection*{Synthesis and characterization of charged colloidal dumplings}

The dumpling colloidal particles {\color{black} (Fig. \ref{fig:expfig1}a)} were prepared using the hydrothermal synthesis method as reported earlier \cite{Wu:2014}. The chemical ingredients used for synthesis, ytterbium nitrate hexahydrate [\ce{Yb(NO3)3} \ce{6H2O}], yttrium nitrate hexahydrate [\ce{Y(NO3)3} \ce{6H2O}], erbium nitrate pentahydrate [\ce{Er(NO3)3} \ce{5H2O}], and sodium fluoride (\ce{NaF})  were all purchased from Sigma Aldrich. Sodium hydroxide (\ce{NaOH}) was purchased from Alfa Aesar. Octanoic acid (OA) was purchased from Acros Organics. In a typical synthesis, 130 mg of \ce{Y(NO3)3}, 40 mg of \ce{Yb(NO3)3}, and 10 mg of \ce{Er(NO3)3} were mixed with 10 ml of deionised water and 13.5 ml of ethanol. After forming a clear transparent solution, 0.35g of \ce{NaOH} and 1.83 g of OA were added into the above solution and stirred continuously at 50$^{\circ}$C for 30 min. Then, 9 ml of 0.2 M \ce{NaF} solution in deionised water was added drop wise to the above solution and stirred continuously for 30 min at 50$^{\circ}$C. The mixture was transferred to a 40 ml Teflon-lined autoclave and kept in an oven for 200$^{\circ}$C for 7 hours. After the reaction, the autoclave was allowed to cool down to room temperature naturally. The particles precipitated at the bottom of the reaction vessel were collected by centrifugation, washed with ethanol and deionised water in sequence, and, lastly, dispersed in 5 ml of cyclohexane. The reaction yields OA functionalised dumpling shaped particles with an average size of 1 $\mu$m, as demonstrated by the scanning electron micrograph of the particles deposited onto a silicon substrate {\color{black} (Fig. \ref{fig:expfig1}a)}. To induce positive surface charges, the particles were treated with an acidic solution. Briefly, 200 $\mu$l of concentrated hydrochloric acid (\ce{HCl}) was mixed with 5 ml of deionised water, and 2.5 ml of the particle solution in cyclohexane was added drop wise to the acid solution and stirred continuously for 12 hours. During this reaction, the OA molecules originally attached to the particle’s surface got detached leaving the particle positively charged. The uncapped particles were collected by the centrifugation and subsequently dispersed in 2.5 ml of ethanol for further use. To prepare the colloidal dispersions, the particle solution was mixed with 5CB (Frinton Laboratories), followed by solvent evaporation at 70$^{\circ}$C for 2 hours and cooling to nematic phase under vigorous mechanical agitation. The nematic dispersions were infiltrated to a glass cell by capillary action and sealed using a fast-setting epoxy. We used planar nematic cells with tangential boundary conditions for experimental observations and video tracking of anisotropic pair interactions of charged and uncharged particles in nematics. To promote unidirectional tangential boundary conditions at the substrate interface, the inner surfaces of the glass substrates were coated with polyvinyl alcohol and then rubbed unidirectionally.

We used an experimental setup built around an inverted Olympus IX81 microscope and a 100{$\times$} (numerical aperture 1.4) oil objective to perform bright-field and polarising microscopy observations. Translational and rotational motion of colloidal particles was recorded with a charge-coupled device camera (Flea, PointGrey) at a rate of 15 frames per second, and the exact $x$-$y$ position {\color{black} fig. S8 and movie S1) } of the dumpling particles as a function of time was then determined from captured sequences of images using motion tracking plugins of ImageJ software. Optical manipulation of dumpling particles was realised with a holographic optical trapping system operating at a wavelength of 1064 nm and integrated with our optical microscopy system. To measure the Debye screening length of the 5CB samples, we used impedance spectroscopy with the Schlumberger SI 1260 impedance analyser.
\\

\subsection*{Calculation of electrostatic potential}

To calculate the electrostatic potential in an anisotropic dielectric medium, we start from Gauss' law, which is given inside a spherical particle with radius $a$ due to the absence of free charges by
\begin{equation}
\nabla\cdot{\bf D}({\bf r})=0, \quad (r<a),
\end{equation}
with ${\bf D}({\bf r})=\epsilon_0\epsilon_\mathrm{p}{\bf E}({\bf r})$ as the displacement field expressed in terms of the vacuum permittivity $\epsilon_0$, particle dielectric constant $\epsilon_\mathrm{p}$, and electric field ${\bf E}({\bf r})$. Outside the particle, in the nematic, we have ions with number densities $\rho_\pm({\bf r})$, and hence the Gauss law reads (in SI units),
\begin{equation}
\nabla\cdot{\bf D}({\bf r})=e[\rho_+({\bf r})-\rho_-({\bf r})], \quad (r>a),
\end{equation}
where now, ${\bf D}({\bf r})=\epsilon_0\boldsymbol{\epsilon}\cdot{\bf E}({\bf r})$, with $\boldsymbol{\epsilon}({\bf r})$ as the (symmetric) dielectric tensor. We can express the electric field in terms of the electrostatic potential $k_\mathrm{B}T\phi({\bf r})/e$, and using the result that within the mean-field approximation the ion densities are Boltzmann distributed, we find Eqs. \eqref{eq:PB} and \eqref{eq:laplace} of the main text, solved numerically with COMSOL Multiphysics under the assumption of a constant charge density $e\sigma$. In the Supplementary Materials, we derive Eqs. \eqref{eq:PB} and \eqref{eq:laplace} also from a free-energy approach.

For $|\phi({\bf r})|\ll 1$ and a constant dielectric tensor, Eq. \eqref{eq:PB} simplifies to
\begin{gather}
\nabla^2\phi({\bf r})=0, \quad (r<a), \\
\Big[(\epsilon_{ij}/\bar{\epsilon}) \partial_i\partial_j-\kappa^2\Big]\phi({\bf r})=0, \quad (r>a),
\end{gather}
to be matched by the {\color{black} constant-charge} boundary condition
\begin{equation}
\hat{\nu}_i\Big[\epsilon_{ij}\partial_j\phi({\bf r})|_{r=a^+}-\epsilon_\mathrm{p}\partial_i\phi({\bf r})|_{r=a^-}\Big]/\bar{\epsilon}=4\pi\lambda_\mathrm{B}^\mathrm{I}\sigma,
\end{equation}
with $\hat{\boldsymbol{\nu}}$ being an outward pointing unit normal vector.
To obtain analytical solutionsn is, however, difficult because of the different symmetry inside the particle compared with outside the particle, which prevents us to find a closed-form expression for $\phi({\bf r})$ while satisfying the constant-charge Neumann boundary condition. We can, however, find a very accurate analytical approximation used in the main text, for which we will sketch the approach here and leave the details of the calculations for the Supplementary Materials.

The most important step in finding an analytical solution is to observe that an approximate solution can be found by solving the auxiliary problem of an ion-penetrable charged shell with surface charge density $e\sigma_\mathrm{S}$ and radius $R$

\begin{equation}
\Big[\left(\epsilon_{ij}/\bar{\epsilon}\right)\partial_i \partial_j-\kappa^2\Big]\varphi({\bf r})=-4\pi\lambda^I_B\sigma_\mathrm{S}\delta(r-R),
\end{equation}
where $\sigma_\mathrm{S}$ and $R$ need to be fitted to the numerically obtained electrostatic potential. 
Then, it turns out that
\begin{equation}
\phi({\bf r})|_{r\geq a}\approx\varphi({\bf r})|_{r\geq a}.
\end{equation}
In general, and especially at high salt concentrations, $R\neq a$ and $\sigma_\mathrm{S}\neq\sigma$, except in the limit where $\kappa\rightarrow 0$. The advantage of the auxiliary problem is that the solution has a closed-form expression with only a double integral left to be evaluated, in terms of the analytically known anisotropic Debye-Hückel (DH) Green's function $G({\bf r},{\bf r}')$, defined by
\begin{equation}
\Big[(\epsilon_{ij}/\bar{\epsilon}) \partial_i\partial_j-\kappa^2\Big]G({\bf r}-{\bf r}')=-4\pi\delta({\bf r}-{\bf r}').
\end{equation}
Hence, we only have to determine the parameters $\gamma=R/a$ and $\alpha=\sigma_\mathrm{S}/\sigma$ based on a two-parameter fit of the numerically obtained \emph{surface} potential, to get the electrostatic potential \emph{everywhere} outside of the particle. The resulting integral expression of $\varphi({\bf r})$ turns out to be almost indistinguishable from the real $\phi({\bf r})$, see the Supplementary Materials for comparative figures. Evaluating a double integral is computationally less expensive than solving the (linearised) PB equation, but the real value of the integral representation comes when calculating pair interactions (see next subsection).

Moreover, the integral expression gives analytical insight. It is now possible to perform a multipole expansion because we have an integral representation of the electrostatic potential outside the particle, as a convolution of a singular charge distribution with the anisotropic DH Green's function. Performing this expansion, the remaining double integrals can be evaluated to find that the decay length is given by Eq. \eqref{eq:anisoscreen}. As is common with multipole expansions, they fail at short distances from the particle, as can be seen from the comparative figures supplied in the Supplementary Materials, but still capture the correct angle dependence for sufficiently large distances. 

\subsection*{Calculation of the screened electrostatic pair interaction potential}
Within linear screening theory $|\phi({\bf r})|\ll1$, ion entropy terms do not contribute to the effective pair potential, and hence the electrostatic part of the pair potential is given by
\begin{equation}
\frac{\Phi_\mathrm{E}}{k_\mathrm{B}T}=\frac{1}{2}\int d{\bf r}\, q({\bf r})\phi({\bf r})-2\frac{U^\text{self}}{k_\mathrm{B}T}, 
\label{eq:phiEE}
\end{equation}
with $q({\bf r})=\sum_{i=1,2}\sigma\delta(|{\bf r}-{\bf R}_i|-a)$, with ${\bf R}_1$ and ${\bf R}_2$ as the center-of-mass position of particle 1 and 2, respectively, and $U_\text{self}$ is the self energy of a single particle. Now $\phi({\bf r})$ is the dimensionless electrostatic potential of the two-body problem. Using the LSA, which entails that the two-body electrostatic potential is given by the sum of the single-particle electrostatic potentials of the two particles, gives the DLVO expression Eq. \eqref{eq:DLVO} if one uses the stress-tensor method \cite{Everts:2020b}. However, applying the LSA directly to Eq. \eqref{eq:phiEE} gives the wrong result because it inappropriately accounts for ion exclusion from the hard core of one particle caused by the double layer of the other particle, which is a curious peculiarity of the free-energy energy route to pair interactions in the theory of charged colloids \cite{Trizac:2002}.

Using the spherical shell renormalisation method, on the other hand, we find that the effective pair potential can be approximated as
\begin{equation}
\frac{\Phi_\mathrm{E}}{k_\mathrm{B}T}\approx\int d{\bf r}\, q_\mathrm{S}({\bf r})\varphi_{2\mathrm{S}}({\bf r}) -2\frac{U_\mathrm{S}^\text{self}}{k_\mathrm{B}T}.
\label{eq:supadupashell}
\end{equation}
Here $q_\mathrm{S}({\bf r})=\sum_{i=1,2}\sigma_\mathrm{S}\delta(|{\bf r}-{\bf R}_i|-R)$ is the charge distribution of two ion-impenetrable charged shells, with the same center-of-mass positions as the spherical particles. Note that singular-charge distributions have an infinite self-energy $U_\mathrm{S}^\text{self}$ that we have to subtract in Eq. \eqref{eq:supadupashell} by using an appropriate regularisation procedure. For example, one way is by giving the shells a finite thickness and taking the thickness in the final step of the calculation to zero. It can straightforwardly be shown that the two-shell electrostatic potential is simply $\varphi_{2\mathrm{S}}({\bf r})=\sum_{i=1,2}\varphi({\bf r}-{\bf R}_i)$. Therefore, the real benefit of this method is  that Eq. \eqref{eq:supadupashell} is determined by the same $\alpha$ and $\gamma$ that are determined from the \emph{single-particle} problem. This method resembles how the electrostatic part of the DLVO expression can be obtained by solving the auxiliary problem of a spherical shell or a point charge, accounting properly for ion-hard core exclusions, although still using the free energy route. The point/shell charge value together with the ions contained within $r<a$ equals the charge on the particle, see, for a more detailed discussion the appendix in Ref. \cite{Trizac:2002} for the shell calculation and Ref.\cite{Everts:2020b} for the point-charge method. Unfortunately, a similar method to determine $\sigma_\mathrm{S}$ does not apply here since $\phi({\bf r})$ is inhomogeneous for $r<a$ when the particle is dispersed in an anisotropic dielectric medium, which therefore requires a numerical fit. However, using the shell method does give the \emph{correct} expression within LSA via a free-energy route.

 Furthermore, the DLVO theory is just the first-order term in a complicated series expansion, where the higher-order terms become more important in the case of high salt concentrations and strong double-layer overlap beyond the LSA \cite{VerweyOverbeek}. We can therefore expect that Eq. \eqref{eq:supadupashell} becomes progressively more inaccurate at high salt concentrations and strong double layer overlaps \cite{Carnie:1993}. Comparing with numerical calculations, we show in the Supplementary Materials that this is indeed the case.

To evaluate Eq. \eqref{eq:supadupashell}, one has to perform four integrals numerically, which is computationally less expensive than solving the (three-dimensional) PB equation, followed by a stress tensor or free-energy integration. To progress further, one can perform again a multipole expansion, but this time of the pair potential from the shell method to obtain an analytically tractable expression that gives more insight in the physics of the effective pair interaction in anisotropic media. See the main text Eq. \eqref{eq:supadupa} and the Supplementary Materials for the derivation of the multipole expansion, as well as comparisons with numerical calculations of the full non-linear theory (but at low charges). Second, we choose to use the multipolar expansion in the main text because we want to treat the screened electrostatic interaction on the same footing as the elastic interaction, which is given on the level of a multipolar expansion as well.

\section*{Supplementary Materials}

\noindent figs. S1 to S7. Various comparative figures of numerically obtained electrostatic potential/effective pair potential with that of the spherical-shell approximation and accompanied multipole expansion. \\
fig. S8. Diffusion of a single charged colloidal dumpling in a nematic liquid crystal. \\
fig. S9. Elastic pair interactions of uncharged dumpling particles. \\
fig. S10. Pair interactions of weakly charged dumpling particles in a nematic LC. \\
movie S1. {\color{black} Brownian motion of a charged dumpling colloidal particle in a nematic liquid crystal. } \\
movie S2. {\color{black} Pair interactions of charged ($\sim300e$) dumpling colloidal particles. } \\
movie S3. {\color{black} Pair interactions of uncharged dumpling colloidal particles. } \\

\makeatletter
\renewcommand\@cite[2]{%
Ref.~#1\ifthenelse{\boolean{@tempswa}}
{, \nolinebreak[3] #2}{}
}
\renewcommand\@biblabel[1]{#1.}
\makeatother
\bibliography{literature1} 

\begin{thebibliography}{57}%
\makeatletter
\providecommand \@ifxundefined [1]{%
 \@ifx{#1\undefined}
}%
\providecommand \@ifnum [1]{%
 \ifnum #1\expandafter \@firstoftwo
 \else \expandafter \@secondoftwo
 \fi
}%
\providecommand \@ifx [1]{%
 \ifx #1\expandafter \@firstoftwo
 \else \expandafter \@secondoftwo
 \fi
}%
\providecommand \natexlab [1]{#1}%
\providecommand \enquote  [1]{``#1''}%
\providecommand \bibnamefont  [1]{#1}%
\providecommand \bibfnamefont [1]{#1}%
\providecommand \citenamefont [1]{#1}%
\providecommand \href@noop [0]{\@secondoftwo}%
\providecommand \href [0]{\begingroup \@sanitize@url \@href}%
\providecommand \@href[1]{\@@startlink{#1}\@@href}%
\providecommand \@@href[1]{\endgroup#1\@@endlink}%
\providecommand \@sanitize@url [0]{\catcode `\\12\catcode `\$12\catcode
  `\&12\catcode `\#12\catcode `\^12\catcode `\_12\catcode `\%12\relax}%
\providecommand \@@startlink[1]{}%
\providecommand \@@endlink[0]{}%
\providecommand \url  [0]{\begingroup\@sanitize@url \@url }%
\providecommand \@url [1]{\endgroup\@href {#1}{\urlprefix }}%
\providecommand \urlprefix  [0]{URL }%
\providecommand \Eprint [0]{\href }%
\providecommand \doibase [0]{https://doi.org/}%
\providecommand \selectlanguage [0]{\@gobble}%
\providecommand \bibinfo  [0]{\@secondoftwo}%
\providecommand \bibfield  [0]{\@secondoftwo}%
\providecommand \translation [1]{[#1]}%
\providecommand \BibitemOpen [0]{}%
\providecommand \bibitemStop [0]{}%
\providecommand \bibitemNoStop [0]{.\EOS\space}%
\providecommand \EOS [0]{\spacefactor3000\relax}%
\providecommand \BibitemShut  [1]{\csname bibitem#1\endcsname}%
\let\auto@bib@innerbib\@empty
\bibitem [{\citenamefont {Derjaguin}\ and\ \citenamefont
  {Landau}(1941)}]{Derjaguin:1948}%
  \BibitemOpen
  \bibfield  {author} {\bibinfo {author} {\bibfnamefont {B.}~\bibnamefont
  {Derjaguin}}\ and\ \bibinfo {author} {\bibfnamefont {L.}~\bibnamefont
  {Landau}},\ }\bibfield  {title} {\bibinfo {title} {Theory of the stability of
  strongly charged lyophobic sols and of the adhesion of strongly charged
  particles in solutions of electrolytes},\ }\href
  {https://doi.org/10.1016/0079-6816(93)90013-L} {\bibfield  {journal}
  {\bibinfo  {journal} {Acta Physicochim. URSS}\ }\textbf {\bibinfo {volume}
  {14}},\ \bibinfo {pages} {633} (\bibinfo {year} {1941})}\BibitemShut
  {NoStop}%
\bibitem [{\citenamefont {Verwey}\ and\ \citenamefont
  {Overbeek}(1948)}]{VerweyOverbeek}%
  \BibitemOpen
  \bibfield  {author} {\bibinfo {author} {\bibfnamefont {E.~J.~W.}\
  \bibnamefont {Verwey}}\ and\ \bibinfo {author} {\bibfnamefont {J.~T.~G.}\
  \bibnamefont {Overbeek}},\ }in\ \href@noop {} {\emph {\bibinfo {booktitle}
  {Theory of the Stability of Lyophobic Colloids}}}\ (\bibinfo  {publisher}
  {Elsevier, New York},\ \bibinfo {year} {1948})\BibitemShut {NoStop}%
\bibitem [{\citenamefont {Vérétout}\ \emph {et~al.}(1989)\citenamefont
  {Vérétout}, \citenamefont {Delaye},\ and\ \citenamefont
  {Tardieu}}]{Veretout:1989}%
  \BibitemOpen
  \bibfield  {author} {\bibinfo {author} {\bibfnamefont {F.}~\bibnamefont
  {Vérétout}}, \bibinfo {author} {\bibfnamefont {M.}~\bibnamefont {Delaye}},\
  and\ \bibinfo {author} {\bibfnamefont {A.}~\bibnamefont {Tardieu}},\
  }\bibfield  {title} {\bibinfo {title} {Molecular basis of eye lens
  transparency: Osmotic pressure and x-ray analysis of crystallin solutions},\
  }\href {https://doi.org/https://doi.org/10.1016/0022-2836(89)90316-1}
  {\bibfield  {journal} {\bibinfo  {journal} {J. Mol. Biol.}\ }\textbf
  {\bibinfo {volume} {205}},\ \bibinfo {pages} {713 } (\bibinfo {year}
  {1989})}\BibitemShut {NoStop}%
\bibitem [{\citenamefont {Ducker}\ \emph {et~al.}(1991)\citenamefont {Ducker},
  \citenamefont {Senden},\ and\ \citenamefont {Pashley}}]{Ducker:1991}%
  \BibitemOpen
  \bibfield  {author} {\bibinfo {author} {\bibfnamefont {W.~A.}\ \bibnamefont
  {Ducker}}, \bibinfo {author} {\bibfnamefont {T.~J.}\ \bibnamefont {Senden}},\
  and\ \bibinfo {author} {\bibfnamefont {R.~M.}\ \bibnamefont {Pashley}},\
  }\bibfield  {title} {\bibinfo {title} {Direct measurement of colloidal forces
  using an atomic force microscope},\ }\href {https://doi.org/10.1038/353239a0}
  {\bibfield  {journal} {\bibinfo  {journal} {Nature}\ }\textbf {\bibinfo
  {volume} {353}},\ \bibinfo {pages} {239} (\bibinfo {year}
  {1991})}\BibitemShut {NoStop}%
\bibitem [{\citenamefont {Carnie}\ and\ \citenamefont
  {Chan}(1993)}]{Carnie:1993}%
  \BibitemOpen
  \bibfield  {author} {\bibinfo {author} {\bibfnamefont {S.~L.}\ \bibnamefont
  {Carnie}}\ and\ \bibinfo {author} {\bibfnamefont {D.~Y.}\ \bibnamefont
  {Chan}},\ }\bibfield  {title} {\bibinfo {title} {Interaction free energy
  between identical spherical colloidal particles: The linearized
  {P}oisson-{B}oltzmann theory},\ }\href
  {https://doi.org/https://doi.org/10.1006/jcis.1993.1039} {\bibfield
  {journal} {\bibinfo  {journal} {J. Colloid Interface Sci}\ }\textbf {\bibinfo
  {volume} {155}},\ \bibinfo {pages} {297 } (\bibinfo {year}
  {1993})}\BibitemShut {NoStop}%
\bibitem [{\citenamefont {Crocker}\ and\ \citenamefont
  {Grier}(1994)}]{Grier:1994}%
  \BibitemOpen
  \bibfield  {author} {\bibinfo {author} {\bibfnamefont {J.~C.}\ \bibnamefont
  {Crocker}}\ and\ \bibinfo {author} {\bibfnamefont {D.~G.}\ \bibnamefont
  {Grier}},\ }\bibfield  {title} {\bibinfo {title} {Microscopic measurement of
  the pair interaction potential of charge-stabilized colloid},\ }\href
  {https://doi.org/10.1103/PhysRevLett.73.352} {\bibfield  {journal} {\bibinfo
  {journal} {Phys. Rev. Lett.}\ }\textbf {\bibinfo {volume} {73}},\ \bibinfo
  {pages} {352} (\bibinfo {year} {1994})}\BibitemShut {NoStop}%
\bibitem [{\citenamefont {Cao}\ \emph {et~al.}(2018)\citenamefont {Cao},
  \citenamefont {Trefalt},\ and\ \citenamefont {Borkovec}}]{Borkovec:2018}%
  \BibitemOpen
  \bibfield  {author} {\bibinfo {author} {\bibfnamefont {T.}~\bibnamefont
  {Cao}}, \bibinfo {author} {\bibfnamefont {G.}~\bibnamefont {Trefalt}},\ and\
  \bibinfo {author} {\bibfnamefont {M.}~\bibnamefont {Borkovec}},\ }\bibfield
  {title} {\bibinfo {title} {Aggregation of colloidal particles in the presence
  of hydrophobic anions: Importance of attractive non-dlvo forces},\ }\href
  {https://doi.org/10.1021/acs.langmuir.8b03191} {\bibfield  {journal}
  {\bibinfo  {journal} {Langmuir}\ }\textbf {\bibinfo {volume} {34}},\ \bibinfo
  {pages} {14368} (\bibinfo {year} {2018})}\BibitemShut {NoStop}%
\bibitem [{\citenamefont {Zhang}\ \emph {et~al.}(2019)\citenamefont {Zhang},
  \citenamefont {Guan}, \citenamefont {Ji}, \citenamefont {Liu}, \citenamefont
  {Jin},\ and\ \citenamefont {Xu}}]{Zhang:2019}%
  \BibitemOpen
  \bibfield  {author} {\bibinfo {author} {\bibfnamefont {M.}~\bibnamefont
  {Zhang}}, \bibinfo {author} {\bibfnamefont {K.}~\bibnamefont {Guan}},
  \bibinfo {author} {\bibfnamefont {Y.}~\bibnamefont {Ji}}, \bibinfo {author}
  {\bibfnamefont {G.}~\bibnamefont {Liu}}, \bibinfo {author} {\bibfnamefont
  {W.}~\bibnamefont {Jin}},\ and\ \bibinfo {author} {\bibfnamefont
  {N.}~\bibnamefont {Xu}},\ }\bibfield  {title} {\bibinfo {title} {Controllable
  ion transport by surface-charged graphene oxide membrane},\ }\href
  {https://doi.org/10.1038/s41467-019-09286-8} {\bibfield  {journal} {\bibinfo
  {journal} {Nat. Commun}\ }\textbf {\bibinfo {volume} {10}},\ \bibinfo {pages}
  {1} (\bibinfo {year} {2019})}\BibitemShut {NoStop}%
\bibitem [{\citenamefont {Dzubiella}\ \emph {et~al.}(2002)\citenamefont
  {Dzubiella}, \citenamefont {Hoffmann},\ and\ \citenamefont
  {L\"owen}}]{Lowen:2002}%
  \BibitemOpen
  \bibfield  {author} {\bibinfo {author} {\bibfnamefont {J.}~\bibnamefont
  {Dzubiella}}, \bibinfo {author} {\bibfnamefont {G.~P.}\ \bibnamefont
  {Hoffmann}},\ and\ \bibinfo {author} {\bibfnamefont {H.}~\bibnamefont
  {L\"owen}},\ }\bibfield  {title} {\bibinfo {title} {Lane formation in
  colloidal mixtures driven by an external field},\ }\href
  {https://doi.org/10.1103/PhysRevE.65.021402} {\bibfield  {journal} {\bibinfo
  {journal} {Phys. Rev. E}\ }\textbf {\bibinfo {volume} {65}},\ \bibinfo
  {pages} {021402} (\bibinfo {year} {2002})}\BibitemShut {NoStop}%
\bibitem [{\citenamefont {Hynninen}\ and\ \citenamefont
  {Dijkstra}(2005)}]{Dijkstra:2005}%
  \BibitemOpen
  \bibfield  {author} {\bibinfo {author} {\bibfnamefont {A.-P.}\ \bibnamefont
  {Hynninen}}\ and\ \bibinfo {author} {\bibfnamefont {M.}~\bibnamefont
  {Dijkstra}},\ }\bibfield  {title} {\bibinfo {title} {Phase diagram of dipolar
  hard and soft spheres: Manipulation of colloidal crystal structures by an
  external field},\ }\href {https://doi.org/10.1103/PhysRevLett.94.138303}
  {\bibfield  {journal} {\bibinfo  {journal} {Phys. Rev. Lett.}\ }\textbf
  {\bibinfo {volume} {94}},\ \bibinfo {pages} {138303} (\bibinfo {year}
  {2005})}\BibitemShut {NoStop}%
\bibitem [{\citenamefont {Zaccone}\ \emph {et~al.}(2009)\citenamefont
  {Zaccone}, \citenamefont {Wu}, \citenamefont {Gentili},\ and\ \citenamefont
  {Morbidelli}}]{Zaccone:2009}%
  \BibitemOpen
  \bibfield  {author} {\bibinfo {author} {\bibfnamefont {A.}~\bibnamefont
  {Zaccone}}, \bibinfo {author} {\bibfnamefont {H.}~\bibnamefont {Wu}},
  \bibinfo {author} {\bibfnamefont {D.}~\bibnamefont {Gentili}},\ and\ \bibinfo
  {author} {\bibfnamefont {M.}~\bibnamefont {Morbidelli}},\ }\bibfield  {title}
  {\bibinfo {title} {Theory of activated-rate processes under shear with
  application to shear-induced aggregation of colloids},\ }\href
  {https://doi.org/10.1103/PhysRevE.80.051404} {\bibfield  {journal} {\bibinfo
  {journal} {Phys. Rev. E}\ }\textbf {\bibinfo {volume} {80}},\ \bibinfo
  {pages} {051404} (\bibinfo {year} {2009})}\BibitemShut {NoStop}%
\bibitem [{\citenamefont {Hopkins}\ \emph {et~al.}(2006)\citenamefont
  {Hopkins}, \citenamefont {Archer},\ and\ \citenamefont {Evans}}]{Evans:2006}%
  \BibitemOpen
  \bibfield  {author} {\bibinfo {author} {\bibfnamefont {P.}~\bibnamefont
  {Hopkins}}, \bibinfo {author} {\bibfnamefont {A.~J.}\ \bibnamefont
  {Archer}},\ and\ \bibinfo {author} {\bibfnamefont {R.}~\bibnamefont
  {Evans}},\ }\bibfield  {title} {\bibinfo {title} {Pair-correlation functions
  and phase separation in a two-component point {Y}ukawa fluid},\ }\href
  {https://doi.org/10.1063/1.2162884} {\bibfield  {journal} {\bibinfo
  {journal} {J. Chem. Phys.}\ }\textbf {\bibinfo {volume} {124}},\ \bibinfo
  {pages} {054503} (\bibinfo {year} {2006})}\BibitemShut {NoStop}%
\bibitem [{\citenamefont {Yoshizawa}\ \emph {et~al.}(2012)\citenamefont
  {Yoshizawa}, \citenamefont {Wakabayashi}, \citenamefont {Yonese},
  \citenamefont {Yamanaka},\ and\ \citenamefont {Royall}}]{Royall:2012}%
  \BibitemOpen
  \bibfield  {author} {\bibinfo {author} {\bibfnamefont {K.}~\bibnamefont
  {Yoshizawa}}, \bibinfo {author} {\bibfnamefont {N.}~\bibnamefont
  {Wakabayashi}}, \bibinfo {author} {\bibfnamefont {M.}~\bibnamefont {Yonese}},
  \bibinfo {author} {\bibfnamefont {J.}~\bibnamefont {Yamanaka}},\ and\
  \bibinfo {author} {\bibfnamefont {C.~P.}\ \bibnamefont {Royall}},\ }\bibfield
   {title} {\bibinfo {title} {Phase separation in binary colloids with charge
  asymmetry},\ }\href {https://doi.org/10.1039/C2SM26164B} {\bibfield
  {journal} {\bibinfo  {journal} {Soft Matter}\ }\textbf {\bibinfo {volume}
  {8}},\ \bibinfo {pages} {11732} (\bibinfo {year} {2012})}\BibitemShut
  {NoStop}%
\bibitem [{\citenamefont {Leunissen}\ \emph {et~al.}(2005)\citenamefont
  {Leunissen}, \citenamefont {Christova}, \citenamefont {Hynninen},
  \citenamefont {Royall}, \citenamefont {Campbell}, \citenamefont {Imhof},
  \citenamefont {Dijkstra}, \citenamefont {Van~Roij},\ and\ \citenamefont
  {Van~Blaaderen}}]{Leunissen:2005}%
  \BibitemOpen
  \bibfield  {author} {\bibinfo {author} {\bibfnamefont {M.~E.}\ \bibnamefont
  {Leunissen}}, \bibinfo {author} {\bibfnamefont {C.~G.}\ \bibnamefont
  {Christova}}, \bibinfo {author} {\bibfnamefont {A.-P.}\ \bibnamefont
  {Hynninen}}, \bibinfo {author} {\bibfnamefont {C.~P.}\ \bibnamefont
  {Royall}}, \bibinfo {author} {\bibfnamefont {A.~I.}\ \bibnamefont
  {Campbell}}, \bibinfo {author} {\bibfnamefont {A.}~\bibnamefont {Imhof}},
  \bibinfo {author} {\bibfnamefont {M.}~\bibnamefont {Dijkstra}}, \bibinfo
  {author} {\bibfnamefont {R.}~\bibnamefont {Van~Roij}},\ and\ \bibinfo
  {author} {\bibfnamefont {A.}~\bibnamefont {Van~Blaaderen}},\ }\bibfield
  {title} {\bibinfo {title} {Ionic colloidal crystals of oppositely charged
  particles},\ }\href {https://doi.org/10.1038/nature03946} {\bibfield
  {journal} {\bibinfo  {journal} {Nature}\ }\textbf {\bibinfo {volume} {437}},\
  \bibinfo {pages} {235} (\bibinfo {year} {2005})}\BibitemShut {NoStop}%
\bibitem [{\citenamefont {Li}\ \emph {et~al.}(2017)\citenamefont {Li},
  \citenamefont {Girard}, \citenamefont {Shen}, \citenamefont {Millan},\ and\
  \citenamefont {Olvera de~la Cruz}}]{Li:2017}%
  \BibitemOpen
  \bibfield  {author} {\bibinfo {author} {\bibfnamefont {Y.}~\bibnamefont
  {Li}}, \bibinfo {author} {\bibfnamefont {M.}~\bibnamefont {Girard}}, \bibinfo
  {author} {\bibfnamefont {M.}~\bibnamefont {Shen}}, \bibinfo {author}
  {\bibfnamefont {J.~A.}\ \bibnamefont {Millan}},\ and\ \bibinfo {author}
  {\bibfnamefont {M.}~\bibnamefont {Olvera de~la Cruz}},\ }\bibfield  {title}
  {\bibinfo {title} {Strong attractions and repulsions mediated by monovalent
  salts},\ }\href {https://doi.org/10.1073/pnas.1713168114} {\bibfield
  {journal} {\bibinfo  {journal} {Proc. Natl. Acad. Sci. U.S.A.}\ }\textbf
  {\bibinfo {volume} {114}},\ \bibinfo {pages} {11838} (\bibinfo {year}
  {2017})}\BibitemShut {NoStop}%
\bibitem [{\citenamefont {van Roij}\ and\ \citenamefont
  {Hansen}(1997)}]{Roij:1997}%
  \BibitemOpen
  \bibfield  {author} {\bibinfo {author} {\bibfnamefont {R.}~\bibnamefont {van
  Roij}}\ and\ \bibinfo {author} {\bibfnamefont {J.-P.}\ \bibnamefont
  {Hansen}},\ }\bibfield  {title} {\bibinfo {title} {Van der {W}aals--like
  instability in suspensions of mutually repelling charged colloids},\ }\href
  {https://doi.org/10.1103/PhysRevLett.79.3082} {\bibfield  {journal} {\bibinfo
   {journal} {Phys. Rev. Lett.}\ }\textbf {\bibinfo {volume} {79}},\ \bibinfo
  {pages} {3082} (\bibinfo {year} {1997})}\BibitemShut {NoStop}%
\bibitem [{\citenamefont {Trizac}\ and\ \citenamefont
  {Levin}(2004)}]{Levin:2004}%
  \BibitemOpen
  \bibfield  {author} {\bibinfo {author} {\bibfnamefont {E.}~\bibnamefont
  {Trizac}}\ and\ \bibinfo {author} {\bibfnamefont {Y.}~\bibnamefont {Levin}},\
  }\bibfield  {title} {\bibinfo {title} {Renormalized jellium model for
  charge-stabilized colloidal suspensions},\ }\href
  {https://doi.org/10.1103/PhysRevE.69.031403} {\bibfield  {journal} {\bibinfo
  {journal} {Phys. Rev. E}\ }\textbf {\bibinfo {volume} {69}},\ \bibinfo
  {pages} {031403} (\bibinfo {year} {2004})}\BibitemShut {NoStop}%
\bibitem [{\citenamefont {Everts}\ \emph {et~al.}(2016)\citenamefont {Everts},
  \citenamefont {van~der Linden}, \citenamefont {van Blaaderen},\ and\
  \citenamefont {van Roij}}]{Everts:2016}%
  \BibitemOpen
  \bibfield  {author} {\bibinfo {author} {\bibfnamefont {J.~C.}\ \bibnamefont
  {Everts}}, \bibinfo {author} {\bibfnamefont {M.~N.}\ \bibnamefont {van~der
  Linden}}, \bibinfo {author} {\bibfnamefont {A.}~\bibnamefont {van
  Blaaderen}},\ and\ \bibinfo {author} {\bibfnamefont {R.}~\bibnamefont {van
  Roij}},\ }\bibfield  {title} {\bibinfo {title} {Alternating strings and
  clusters in suspensions of charged colloids},\ }\href
  {https://doi.org/10.1039/C6SM01283C} {\bibfield  {journal} {\bibinfo
  {journal} {Soft Matter}\ }\textbf {\bibinfo {volume} {12}},\ \bibinfo {pages}
  {6610} (\bibinfo {year} {2016})}\BibitemShut {NoStop}%
\bibitem [{\citenamefont {Ninham}\ and\ \citenamefont
  {Parsegian}(1971)}]{Ninham:1971}%
  \BibitemOpen
  \bibfield  {author} {\bibinfo {author} {\bibfnamefont {B.~W.}\ \bibnamefont
  {Ninham}}\ and\ \bibinfo {author} {\bibfnamefont {V.}~\bibnamefont
  {Parsegian}},\ }\bibfield  {title} {\bibinfo {title} {Electrostatic potential
  between surfaces bearing ionizable groups in ionic equilibrium with
  physiologic saline solution},\ }\href
  {https://doi.org/10.1016/0022-5193(71)90019-1} {\bibfield  {journal}
  {\bibinfo  {journal} {J. Theor. Biol.}\ }\textbf {\bibinfo {volume} {31}},\
  \bibinfo {pages} {405 } (\bibinfo {year} {1971})}\BibitemShut {NoStop}%
\bibitem [{\citenamefont {Popa}\ \emph {et~al.}(2010)\citenamefont {Popa},
  \citenamefont {Sinha}, \citenamefont {Finessi}, \citenamefont {Maroni},
  \citenamefont {Papastavrou},\ and\ \citenamefont {Borkovec}}]{Borkovec:2010}%
  \BibitemOpen
  \bibfield  {author} {\bibinfo {author} {\bibfnamefont {I.}~\bibnamefont
  {Popa}}, \bibinfo {author} {\bibfnamefont {P.}~\bibnamefont {Sinha}},
  \bibinfo {author} {\bibfnamefont {M.}~\bibnamefont {Finessi}}, \bibinfo
  {author} {\bibfnamefont {P.}~\bibnamefont {Maroni}}, \bibinfo {author}
  {\bibfnamefont {G.}~\bibnamefont {Papastavrou}},\ and\ \bibinfo {author}
  {\bibfnamefont {M.}~\bibnamefont {Borkovec}},\ }\bibfield  {title} {\bibinfo
  {title} {Importance of charge regulation in attractive double-layer forces
  between dissimilar surfaces},\ }\href
  {https://doi.org/10.1103/PhysRevLett.104.228301} {\bibfield  {journal}
  {\bibinfo  {journal} {Phys. Rev. Lett.}\ }\textbf {\bibinfo {volume} {104}},\
  \bibinfo {pages} {228301} (\bibinfo {year} {2010})}\BibitemShut {NoStop}%
\bibitem [{\citenamefont {dos Santos}\ and\ \citenamefont
  {Levin}(2019)}]{Levin:2019}%
  \BibitemOpen
  \bibfield  {author} {\bibinfo {author} {\bibfnamefont {A.~P.}\ \bibnamefont
  {dos Santos}}\ and\ \bibinfo {author} {\bibfnamefont {Y.}~\bibnamefont
  {Levin}},\ }\bibfield  {title} {\bibinfo {title} {Like-charge attraction
  between metal nanoparticles in a $1\ensuremath{\mathbin:}1$ electrolyte
  solution},\ }\href {https://doi.org/10.1103/PhysRevLett.122.248005}
  {\bibfield  {journal} {\bibinfo  {journal} {Phys. Rev. Lett.}\ }\textbf
  {\bibinfo {volume} {122}},\ \bibinfo {pages} {248005} (\bibinfo {year}
  {2019})}\BibitemShut {NoStop}%
\bibitem [{\citenamefont {Alexander}\ \emph {et~al.}(1984)\citenamefont
  {Alexander}, \citenamefont {Chaikin}, \citenamefont {Grant}, \citenamefont
  {Morales}, \citenamefont {Pincus},\ and\ \citenamefont
  {Hone}}]{Alexander:1984}%
  \BibitemOpen
  \bibfield  {author} {\bibinfo {author} {\bibfnamefont {S.}~\bibnamefont
  {Alexander}}, \bibinfo {author} {\bibfnamefont {P.~M.}\ \bibnamefont
  {Chaikin}}, \bibinfo {author} {\bibfnamefont {P.}~\bibnamefont {Grant}},
  \bibinfo {author} {\bibfnamefont {G.~J.}\ \bibnamefont {Morales}}, \bibinfo
  {author} {\bibfnamefont {P.}~\bibnamefont {Pincus}},\ and\ \bibinfo {author}
  {\bibfnamefont {D.}~\bibnamefont {Hone}},\ }\bibfield  {title} {\bibinfo
  {title} {Charge renormalization, osmotic pressure, and bulk modulus of
  colloidal crystals: Theory},\ }\href {https://doi.org/10.1063/1.446600}
  {\bibfield  {journal} {\bibinfo  {journal} {J. Chem. Phys.}\ }\textbf
  {\bibinfo {volume} {80}},\ \bibinfo {pages} {5776} (\bibinfo {year}
  {1984})}\BibitemShut {NoStop}%
\bibitem [{\citenamefont {Naji}\ \emph {et~al.}(2013)\citenamefont {Naji},
  \citenamefont {Kanduč}, \citenamefont {Forsman},\ and\ \citenamefont
  {Podgornik}}]{Naji:2013}%
  \BibitemOpen
  \bibfield  {author} {\bibinfo {author} {\bibfnamefont {A.}~\bibnamefont
  {Naji}}, \bibinfo {author} {\bibfnamefont {M.}~\bibnamefont {Kanduč}},
  \bibinfo {author} {\bibfnamefont {J.}~\bibnamefont {Forsman}},\ and\ \bibinfo
  {author} {\bibfnamefont {R.}~\bibnamefont {Podgornik}},\ }\bibfield  {title}
  {\bibinfo {title} {Perspective: Coulomb fluids—weak coupling, strong
  coupling, in between and beyond},\ }\href {https://doi.org/10.1063/1.4824681}
  {\bibfield  {journal} {\bibinfo  {journal} {J. Chem. Phys.}\ }\textbf
  {\bibinfo {volume} {139}},\ \bibinfo {pages} {150901} (\bibinfo {year}
  {2013})}\BibitemShut {NoStop}%
\bibitem [{\citenamefont {Bostr\"om}\ \emph {et~al.}(2001)\citenamefont
  {Bostr\"om}, \citenamefont {Williams},\ and\ \citenamefont
  {Ninham}}]{Ninham:2001}%
  \BibitemOpen
  \bibfield  {author} {\bibinfo {author} {\bibfnamefont {M.}~\bibnamefont
  {Bostr\"om}}, \bibinfo {author} {\bibfnamefont {D.~R.~M.}\ \bibnamefont
  {Williams}},\ and\ \bibinfo {author} {\bibfnamefont {B.~W.}\ \bibnamefont
  {Ninham}},\ }\bibfield  {title} {\bibinfo {title} {Specific ion effects: Why
  {DLVO} theory fails for biology and colloid systems},\ }\href
  {https://doi.org/10.1103/PhysRevLett.87.168103} {\bibfield  {journal}
  {\bibinfo  {journal} {Phys. Rev. Lett.}\ }\textbf {\bibinfo {volume} {87}},\
  \bibinfo {pages} {168103} (\bibinfo {year} {2001})}\BibitemShut {NoStop}%
\bibitem [{\citenamefont {Silvera~Batista}\ \emph {et~al.}(2015)\citenamefont
  {Silvera~Batista}, \citenamefont {Larson},\ and\ \citenamefont
  {Kotov}}]{Kotov:2015}%
  \BibitemOpen
  \bibfield  {author} {\bibinfo {author} {\bibfnamefont {C.~A.}\ \bibnamefont
  {Silvera~Batista}}, \bibinfo {author} {\bibfnamefont {R.~G.}\ \bibnamefont
  {Larson}},\ and\ \bibinfo {author} {\bibfnamefont {N.~A.}\ \bibnamefont
  {Kotov}},\ }\bibfield  {title} {\bibinfo {title} {Nonadditivity of
  nanoparticle interactions},\ }\bibfield  {journal} {\bibinfo  {journal}
  {Science}\ }\textbf {\bibinfo {volume} {350}},\ \href
  {https://doi.org/10.1126/science.1242477} {10.1126/science.1242477} (\bibinfo
  {year} {2015})\BibitemShut {NoStop}%
\bibitem [{\citenamefont {Gompper}\ \emph {et~al.}(2020)\citenamefont
  {Gompper}, \citenamefont {Winkler}, \citenamefont {Speck}, \citenamefont
  {Solon}, \citenamefont {Nardini}, \citenamefont {Peruani}, \citenamefont
  {Löwen}, \citenamefont {Golestanian}, \citenamefont {Kaupp}, \citenamefont
  {Alvarez}, \citenamefont {Ki{\o}rboe}, \citenamefont {Lauga}, \citenamefont
  {Poon}, \citenamefont {DeSimone}, \citenamefont {Mui{\~{n}}os-Landin},
  \citenamefont {Fischer}, \citenamefont {Söker}, \citenamefont {Cichos},
  \citenamefont {Kapral}, \citenamefont {Gaspard}, \citenamefont {Ripoll},
  \citenamefont {Sagues}, \citenamefont {Doostmohammadi}, \citenamefont
  {Yeomans}, \citenamefont {Aranson}, \citenamefont {Bechinger}, \citenamefont
  {Stark}, \citenamefont {Hemelrijk}, \citenamefont {Nedelec}, \citenamefont
  {Sarkar}, \citenamefont {Aryaksama}, \citenamefont {Lacroix}, \citenamefont
  {Duclos}, \citenamefont {Yashunsky}, \citenamefont {Silberzan}, \citenamefont
  {Arroyo},\ and\ \citenamefont {Kale}}]{Gompper:2020}%
  \BibitemOpen
  \bibfield  {author} {\bibinfo {author} {\bibfnamefont {G.}~\bibnamefont
  {Gompper}}, \bibinfo {author} {\bibfnamefont {R.~G.}\ \bibnamefont
  {Winkler}}, \bibinfo {author} {\bibfnamefont {T.}~\bibnamefont {Speck}},
  \bibinfo {author} {\bibfnamefont {A.}~\bibnamefont {Solon}}, \bibinfo
  {author} {\bibfnamefont {C.}~\bibnamefont {Nardini}}, \bibinfo {author}
  {\bibfnamefont {F.}~\bibnamefont {Peruani}}, \bibinfo {author} {\bibfnamefont
  {H.}~\bibnamefont {Löwen}}, \bibinfo {author} {\bibfnamefont
  {R.}~\bibnamefont {Golestanian}}, \bibinfo {author} {\bibfnamefont {U.~B.}\
  \bibnamefont {Kaupp}}, \bibinfo {author} {\bibfnamefont {L.}~\bibnamefont
  {Alvarez}}, \bibinfo {author} {\bibfnamefont {T.}~\bibnamefont {Ki{\o}rboe}},
  \bibinfo {author} {\bibfnamefont {E.}~\bibnamefont {Lauga}}, \bibinfo
  {author} {\bibfnamefont {W.~C.~K.}\ \bibnamefont {Poon}}, \bibinfo {author}
  {\bibfnamefont {A.}~\bibnamefont {DeSimone}}, \bibinfo {author}
  {\bibfnamefont {S.}~\bibnamefont {Mui{\~{n}}os-Landin}}, \bibinfo {author}
  {\bibfnamefont {A.}~\bibnamefont {Fischer}}, \bibinfo {author} {\bibfnamefont
  {N.~A.}\ \bibnamefont {Söker}}, \bibinfo {author} {\bibfnamefont
  {F.}~\bibnamefont {Cichos}}, \bibinfo {author} {\bibfnamefont
  {R.}~\bibnamefont {Kapral}}, \bibinfo {author} {\bibfnamefont
  {P.}~\bibnamefont {Gaspard}}, \bibinfo {author} {\bibfnamefont
  {M.}~\bibnamefont {Ripoll}}, \bibinfo {author} {\bibfnamefont
  {F.}~\bibnamefont {Sagues}}, \bibinfo {author} {\bibfnamefont
  {A.}~\bibnamefont {Doostmohammadi}}, \bibinfo {author} {\bibfnamefont
  {J.~M.}\ \bibnamefont {Yeomans}}, \bibinfo {author} {\bibfnamefont {I.~S.}\
  \bibnamefont {Aranson}}, \bibinfo {author} {\bibfnamefont {C.}~\bibnamefont
  {Bechinger}}, \bibinfo {author} {\bibfnamefont {H.}~\bibnamefont {Stark}},
  \bibinfo {author} {\bibfnamefont {C.~K.}\ \bibnamefont {Hemelrijk}}, \bibinfo
  {author} {\bibfnamefont {F.~J.}\ \bibnamefont {Nedelec}}, \bibinfo {author}
  {\bibfnamefont {T.}~\bibnamefont {Sarkar}}, \bibinfo {author} {\bibfnamefont
  {T.}~\bibnamefont {Aryaksama}}, \bibinfo {author} {\bibfnamefont
  {M.}~\bibnamefont {Lacroix}}, \bibinfo {author} {\bibfnamefont
  {G.}~\bibnamefont {Duclos}}, \bibinfo {author} {\bibfnamefont
  {V.}~\bibnamefont {Yashunsky}}, \bibinfo {author} {\bibfnamefont
  {P.}~\bibnamefont {Silberzan}}, \bibinfo {author} {\bibfnamefont
  {M.}~\bibnamefont {Arroyo}},\ and\ \bibinfo {author} {\bibfnamefont
  {S.}~\bibnamefont {Kale}},\ }\bibfield  {title} {\bibinfo {title} {The 2020
  motile active matter roadmap},\ }\href
  {https://doi.org/10.1088/1361-648x/ab6348} {\bibfield  {journal} {\bibinfo
  {journal} {J. Phys. Condens. Matter}\ }\textbf {\bibinfo {volume} {32}},\
  \bibinfo {pages} {193001} (\bibinfo {year} {2020})}\BibitemShut {NoStop}%
\bibitem [{\citenamefont {Rowan}\ \emph {et~al.}(2000)\citenamefont {Rowan},
  \citenamefont {Hansen},\ and\ \citenamefont {Trizac}}]{Trizac:2000}%
  \BibitemOpen
  \bibfield  {author} {\bibinfo {author} {\bibfnamefont {D.~G.}\ \bibnamefont
  {Rowan}}, \bibinfo {author} {\bibfnamefont {J.-P.}\ \bibnamefont {Hansen}},\
  and\ \bibinfo {author} {\bibfnamefont {E.}~\bibnamefont {Trizac}},\
  }\bibfield  {title} {\bibinfo {title} {Screened electrostatic interactions
  between clay platelets},\ }\href {https://doi.org/10.1080/002689700417493}
  {\bibfield  {journal} {\bibinfo  {journal} {Mol. Phys.}\ }\textbf {\bibinfo
  {volume} {98}},\ \bibinfo {pages} {1369} (\bibinfo {year}
  {2000})}\BibitemShut {NoStop}%
\bibitem [{\citenamefont {Trizac}\ \emph {et~al.}(2002)\citenamefont {Trizac},
  \citenamefont {Bocquet}, \citenamefont {Agra}, \citenamefont {Weis},\ and\
  \citenamefont {Aubouy}}]{Trizac:2002}%
  \BibitemOpen
  \bibfield  {author} {\bibinfo {author} {\bibfnamefont {E.}~\bibnamefont
  {Trizac}}, \bibinfo {author} {\bibfnamefont {L.}~\bibnamefont {Bocquet}},
  \bibinfo {author} {\bibfnamefont {R.}~\bibnamefont {Agra}}, \bibinfo {author}
  {\bibfnamefont {J.-J.}\ \bibnamefont {Weis}},\ and\ \bibinfo {author}
  {\bibfnamefont {M.}~\bibnamefont {Aubouy}},\ }\bibfield  {title} {\bibinfo
  {title} {Effective interactions and phase behaviour for a model clay
  suspension in an electrolyte},\ }\href
  {https://doi.org/10.1088/0953-8984/14/40/322} {\bibfield  {journal} {\bibinfo
   {journal} {J. Phys.: Cond. Matt.}\ }\textbf {\bibinfo {volume} {14}},\
  \bibinfo {pages} {9339} (\bibinfo {year} {2002})}\BibitemShut {NoStop}%
\bibitem [{\citenamefont {Álvarez}\ and\ \citenamefont
  {Téllez}(2010)}]{Tellez:2010}%
  \BibitemOpen
  \bibfield  {author} {\bibinfo {author} {\bibfnamefont {C.}~\bibnamefont
  {Álvarez}}\ and\ \bibinfo {author} {\bibfnamefont {G.}~\bibnamefont
  {Téllez}},\ }\bibfield  {title} {\bibinfo {title} {Screening of charged
  spheroidal colloidal particles},\ }\href {https://doi.org/10.1063/1.3486558}
  {\bibfield  {journal} {\bibinfo  {journal} {J. Chem. Phys.}\ }\textbf
  {\bibinfo {volume} {133}},\ \bibinfo {pages} {144908} (\bibinfo {year}
  {2010})}\BibitemShut {NoStop}%
\bibitem [{\citenamefont {Bonthuis}\ \emph {et~al.}(2011)\citenamefont
  {Bonthuis}, \citenamefont {Gekle},\ and\ \citenamefont {Netz}}]{Netz:2011}%
  \BibitemOpen
  \bibfield  {author} {\bibinfo {author} {\bibfnamefont {D.~J.}\ \bibnamefont
  {Bonthuis}}, \bibinfo {author} {\bibfnamefont {S.}~\bibnamefont {Gekle}},\
  and\ \bibinfo {author} {\bibfnamefont {R.~R.}\ \bibnamefont {Netz}},\
  }\bibfield  {title} {\bibinfo {title} {Dielectric profile of interfacial
  water and its effect on double-layer capacitance},\ }\href
  {https://doi.org/10.1103/PhysRevLett.107.166102} {\bibfield  {journal}
  {\bibinfo  {journal} {Phys. Rev. Lett.}\ }\textbf {\bibinfo {volume} {107}},\
  \bibinfo {pages} {166102} (\bibinfo {year} {2011})}\BibitemShut {NoStop}%
\bibitem [{\citenamefont {Loche}\ \emph {et~al.}(2020)\citenamefont {Loche},
  \citenamefont {Ayaz}, \citenamefont {Wolde-Kidan}, \citenamefont {Schlaich},\
  and\ \citenamefont {Netz}}]{Loche:2020}%
  \BibitemOpen
  \bibfield  {author} {\bibinfo {author} {\bibfnamefont {P.}~\bibnamefont
  {Loche}}, \bibinfo {author} {\bibfnamefont {C.}~\bibnamefont {Ayaz}},
  \bibinfo {author} {\bibfnamefont {A.}~\bibnamefont {Wolde-Kidan}}, \bibinfo
  {author} {\bibfnamefont {A.}~\bibnamefont {Schlaich}},\ and\ \bibinfo
  {author} {\bibfnamefont {R.~R.}\ \bibnamefont {Netz}},\ }\bibfield  {title}
  {\bibinfo {title} {Universal and nonuniversal aspects of electrostatics in
  aqueous nanoconfinement},\ }\href {https://doi.org/10.1021/acs.jpcb.0c01967}
  {\bibfield  {journal} {\bibinfo  {journal} {J. Phys. Chem. B}\ }\textbf
  {\bibinfo {volume} {124}},\ \bibinfo {pages} {4365} (\bibinfo {year}
  {2020})}\BibitemShut {NoStop}%
\bibitem [{\citenamefont {Poulin}\ \emph {et~al.}(1997)\citenamefont {Poulin},
  \citenamefont {Stark}, \citenamefont {Lubensky},\ and\ \citenamefont
  {Weitz}}]{Poulin:1997}%
  \BibitemOpen
  \bibfield  {author} {\bibinfo {author} {\bibfnamefont {P.}~\bibnamefont
  {Poulin}}, \bibinfo {author} {\bibfnamefont {H.}~\bibnamefont {Stark}},
  \bibinfo {author} {\bibfnamefont {T.~C.}\ \bibnamefont {Lubensky}},\ and\
  \bibinfo {author} {\bibfnamefont {D.~A.}\ \bibnamefont {Weitz}},\ }\bibfield
  {title} {\bibinfo {title} {Novel colloidal interactions in anisotropic
  fluids},\ }\href {https://doi.org/10.1126/science.275.5307.1770} {\bibfield
  {journal} {\bibinfo  {journal} {Science}\ }\textbf {\bibinfo {volume}
  {275}},\ \bibinfo {pages} {1770} (\bibinfo {year} {1997})}\BibitemShut
  {NoStop}%
\bibitem [{\citenamefont {Muševič}\ \emph {et~al.}(2006)\citenamefont
  {Muševič}, \citenamefont {Škarabot}, \citenamefont {Tkalec}, \citenamefont
  {Ravnik},\ and\ \citenamefont {Žumer}}]{Musevic:2006}%
  \BibitemOpen
  \bibfield  {author} {\bibinfo {author} {\bibfnamefont {I.}~\bibnamefont
  {Muševič}}, \bibinfo {author} {\bibfnamefont {M.}~\bibnamefont
  {Škarabot}}, \bibinfo {author} {\bibfnamefont {U.}~\bibnamefont {Tkalec}},
  \bibinfo {author} {\bibfnamefont {M.}~\bibnamefont {Ravnik}},\ and\ \bibinfo
  {author} {\bibfnamefont {S.}~\bibnamefont {Žumer}},\ }\bibfield  {title}
  {\bibinfo {title} {Two-dimensional nematic colloidal crystals self-assembled
  by topological defects},\ }\href {https://doi.org/10.1126/science.1129660}
  {\bibfield  {journal} {\bibinfo  {journal} {Science}\ }\textbf {\bibinfo
  {volume} {313}},\ \bibinfo {pages} {954} (\bibinfo {year}
  {2006})}\BibitemShut {NoStop}%
\bibitem [{\citenamefont {Lapointe}\ \emph {et~al.}(2009)\citenamefont
  {Lapointe}, \citenamefont {Mason},\ and\ \citenamefont
  {Smalyukh}}]{Lapointe:2009}%
  \BibitemOpen
  \bibfield  {author} {\bibinfo {author} {\bibfnamefont {C.~P.}\ \bibnamefont
  {Lapointe}}, \bibinfo {author} {\bibfnamefont {T.~G.}\ \bibnamefont
  {Mason}},\ and\ \bibinfo {author} {\bibfnamefont {I.~I.}\ \bibnamefont
  {Smalyukh}},\ }\bibfield  {title} {\bibinfo {title} {Shape-controlled
  colloidal interactions in nematic liquid crystals},\ }\href
  {https://doi.org/10.1126/science.1176587} {\bibfield  {journal} {\bibinfo
  {journal} {Science}\ }\textbf {\bibinfo {volume} {326}},\ \bibinfo {pages}
  {1083} (\bibinfo {year} {2009})}\BibitemShut {NoStop}%
\bibitem [{\citenamefont {Cavallaro}\ \emph {et~al.}(2013)\citenamefont
  {Cavallaro}, \citenamefont {Gharbi}, \citenamefont {Beller}, \citenamefont
  {{\v C}opar}, \citenamefont {Shi}, \citenamefont {Baumgart}, \citenamefont
  {Yang}, \citenamefont {Kamien},\ and\ \citenamefont {Stebe}}]{Stebe:2013}%
  \BibitemOpen
  \bibfield  {author} {\bibinfo {author} {\bibfnamefont {M.}~\bibnamefont
  {Cavallaro}}, \bibinfo {author} {\bibfnamefont {M.~A.}\ \bibnamefont
  {Gharbi}}, \bibinfo {author} {\bibfnamefont {D.~A.}\ \bibnamefont {Beller}},
  \bibinfo {author} {\bibfnamefont {S.}~\bibnamefont {{\v C}opar}}, \bibinfo
  {author} {\bibfnamefont {Z.}~\bibnamefont {Shi}}, \bibinfo {author}
  {\bibfnamefont {T.}~\bibnamefont {Baumgart}}, \bibinfo {author}
  {\bibfnamefont {S.}~\bibnamefont {Yang}}, \bibinfo {author} {\bibfnamefont
  {R.~D.}\ \bibnamefont {Kamien}},\ and\ \bibinfo {author} {\bibfnamefont
  {K.~J.}\ \bibnamefont {Stebe}},\ }\bibfield  {title} {\bibinfo {title}
  {Exploiting imperfections in the bulk to direct assembly of surface
  colloids},\ }\href {https://doi.org/10.1073/pnas.1313551110} {\bibfield
  {journal} {\bibinfo  {journal} {Proc. Natl. Acad. Sci. U.S.A}\ }\textbf
  {\bibinfo {volume} {110}},\ \bibinfo {pages} {18804} (\bibinfo {year}
  {2013})}\BibitemShut {NoStop}%
\bibitem [{\citenamefont {Yuan}\ and\ \citenamefont
  {Smalyukh}(2015)}]{Yuan:2015}%
  \BibitemOpen
  \bibfield  {author} {\bibinfo {author} {\bibfnamefont {Y.}~\bibnamefont
  {Yuan}}\ and\ \bibinfo {author} {\bibfnamefont {I.~I.}\ \bibnamefont
  {Smalyukh}},\ }\bibfield  {title} {\bibinfo {title} {Topological nanocolloids
  with facile electric switching of plasmonic properties},\ }\href
  {https://doi.org/10.1364/OL.40.005630} {\bibfield  {journal} {\bibinfo
  {journal} {Opt. Lett.}\ }\textbf {\bibinfo {volume} {40}},\ \bibinfo {pages}
  {5630} (\bibinfo {year} {2015})}\BibitemShut {NoStop}%
\bibitem [{\citenamefont {Lubensky}\ \emph {et~al.}(1998)\citenamefont
  {Lubensky}, \citenamefont {Pettey}, \citenamefont {Currier},\ and\
  \citenamefont {Stark}}]{Lubensky:1998}%
  \BibitemOpen
  \bibfield  {author} {\bibinfo {author} {\bibfnamefont {T.~C.}\ \bibnamefont
  {Lubensky}}, \bibinfo {author} {\bibfnamefont {D.}~\bibnamefont {Pettey}},
  \bibinfo {author} {\bibfnamefont {N.}~\bibnamefont {Currier}},\ and\ \bibinfo
  {author} {\bibfnamefont {H.}~\bibnamefont {Stark}},\ }\bibfield  {title}
  {\bibinfo {title} {Topological defects and interactions in nematic
  emulsions},\ }\href {https://doi.org/10.1103/PhysRevE.57.610} {\bibfield
  {journal} {\bibinfo  {journal} {Phys. Rev. E}\ }\textbf {\bibinfo {volume}
  {57}},\ \bibinfo {pages} {610} (\bibinfo {year} {1998})}\BibitemShut
  {NoStop}%
\bibitem [{\citenamefont {Lev}\ \emph {et~al.}(2002)\citenamefont {Lev},
  \citenamefont {Chernyshuk}, \citenamefont {Tomchuk},\ and\ \citenamefont
  {Yokoyama}}]{Lev:2002}%
  \BibitemOpen
  \bibfield  {author} {\bibinfo {author} {\bibfnamefont {B.~I.}\ \bibnamefont
  {Lev}}, \bibinfo {author} {\bibfnamefont {S.~B.}\ \bibnamefont {Chernyshuk}},
  \bibinfo {author} {\bibfnamefont {P.~M.}\ \bibnamefont {Tomchuk}},\ and\
  \bibinfo {author} {\bibfnamefont {H.}~\bibnamefont {Yokoyama}},\ }\bibfield
  {title} {\bibinfo {title} {Symmetry breaking and interaction of colloidal
  particles in nematic liquid crystals},\ }\href
  {https://doi.org/10.1103/PhysRevE.65.021709} {\bibfield  {journal} {\bibinfo
  {journal} {Phys. Rev. E}\ }\textbf {\bibinfo {volume} {65}},\ \bibinfo
  {pages} {021709} (\bibinfo {year} {2002})}\BibitemShut {NoStop}%
\bibitem [{\citenamefont {Yuan}\ \emph {et~al.}(2020)\citenamefont {Yuan},
  \citenamefont {Tasinkevych},\ and\ \citenamefont {Smalyukh}}]{Yuan:2020}%
  \BibitemOpen
  \bibfield  {author} {\bibinfo {author} {\bibfnamefont {Y.}~\bibnamefont
  {Yuan}}, \bibinfo {author} {\bibfnamefont {M.}~\bibnamefont {Tasinkevych}},\
  and\ \bibinfo {author} {\bibfnamefont {I.~I.}\ \bibnamefont {Smalyukh}},\
  }\bibfield  {title} {\bibinfo {title} {Colloidal interactions and unusual
  crystallization versus de-mixing of elastic multipoles formed by gold
  mesoflowers},\ }\href {https://doi.org/10.1038/s41467-019-14031-2} {\bibfield
   {journal} {\bibinfo  {journal} {Nat. Commun.}\ }\textbf {\bibinfo {volume}
  {11}},\ \bibinfo {pages} {188} (\bibinfo {year} {2020})}\BibitemShut
  {NoStop}%
\bibitem [{\citenamefont {Mundoor}\ \emph {et~al.}(2016)\citenamefont
  {Mundoor}, \citenamefont {Senyuk},\ and\ \citenamefont
  {Smalyukh}}]{Mundoor:2016}%
  \BibitemOpen
  \bibfield  {author} {\bibinfo {author} {\bibfnamefont {H.}~\bibnamefont
  {Mundoor}}, \bibinfo {author} {\bibfnamefont {B.}~\bibnamefont {Senyuk}},\
  and\ \bibinfo {author} {\bibfnamefont {I.~I.}\ \bibnamefont {Smalyukh}},\
  }\bibfield  {title} {\bibinfo {title} {Triclinic nematic colloidal crystals
  from competing elastic and electrostatic interactions},\ }\href
  {https://doi.org/10.1126/science.aaf0801} {\bibfield  {journal} {\bibinfo
  {journal} {Science}\ }\textbf {\bibinfo {volume} {352}},\ \bibinfo {pages}
  {69} (\bibinfo {year} {2016})}\BibitemShut {NoStop}%
\bibitem [{\citenamefont {Mundoor}\ \emph {et~al.}(2018)\citenamefont
  {Mundoor}, \citenamefont {Park}, \citenamefont {Senyuk}, \citenamefont
  {Wensink},\ and\ \citenamefont {Smalyukh}}]{Mundoor:2018}%
  \BibitemOpen
  \bibfield  {author} {\bibinfo {author} {\bibfnamefont {H.}~\bibnamefont
  {Mundoor}}, \bibinfo {author} {\bibfnamefont {S.}~\bibnamefont {Park}},
  \bibinfo {author} {\bibfnamefont {B.}~\bibnamefont {Senyuk}}, \bibinfo
  {author} {\bibfnamefont {H.~H.}\ \bibnamefont {Wensink}},\ and\ \bibinfo
  {author} {\bibfnamefont {I.~I.}\ \bibnamefont {Smalyukh}},\ }\bibfield
  {title} {\bibinfo {title} {Hybrid molecular-colloidal liquid crystals},\
  }\href {https://doi.org/10.1126/science.aap9359} {\bibfield  {journal}
  {\bibinfo  {journal} {Science}\ }\textbf {\bibinfo {volume} {360}},\ \bibinfo
  {pages} {768} (\bibinfo {year} {2018})}\BibitemShut {NoStop}%
\bibitem [{\citenamefont {Mundoor}\ \emph {et~al.}(2019)\citenamefont
  {Mundoor}, \citenamefont {Senyuk}, \citenamefont {Almansouri}, \citenamefont
  {Park}, \citenamefont {Fleury},\ and\ \citenamefont
  {Smalyukh}}]{Mundoor:2019}%
  \BibitemOpen
  \bibfield  {author} {\bibinfo {author} {\bibfnamefont {H.}~\bibnamefont
  {Mundoor}}, \bibinfo {author} {\bibfnamefont {B.}~\bibnamefont {Senyuk}},
  \bibinfo {author} {\bibfnamefont {M.}~\bibnamefont {Almansouri}}, \bibinfo
  {author} {\bibfnamefont {S.}~\bibnamefont {Park}}, \bibinfo {author}
  {\bibfnamefont {B.}~\bibnamefont {Fleury}},\ and\ \bibinfo {author}
  {\bibfnamefont {I.~I.}\ \bibnamefont {Smalyukh}},\ }\bibfield  {title}
  {\bibinfo {title} {Electrostatically controlled surface boundary conditions
  in nematic liquid crystals and colloids},\ }\bibfield  {journal} {\bibinfo
  {journal} {Sci. Adv.}\ }\textbf {\bibinfo {volume} {5}},\ \href
  {https://doi.org/10.1126/sciadv.aax4257} {10.1126/sciadv.aax4257} (\bibinfo
  {year} {2019})\BibitemShut {NoStop}%
\bibitem [{\citenamefont {Gu}\ and\ \citenamefont
  {Abbott}(2000)}]{Abbott:2000}%
  \BibitemOpen
  \bibfield  {author} {\bibinfo {author} {\bibfnamefont {Y.}~\bibnamefont
  {Gu}}\ and\ \bibinfo {author} {\bibfnamefont {N.~L.}\ \bibnamefont
  {Abbott}},\ }\bibfield  {title} {\bibinfo {title} {Observation of saturn-ring
  defects around solid microspheres in nematic liquid crystals},\ }\href
  {https://doi.org/10.1103/PhysRevLett.85.4719} {\bibfield  {journal} {\bibinfo
   {journal} {Phys. Rev. Lett.}\ }\textbf {\bibinfo {volume} {85}},\ \bibinfo
  {pages} {4719} (\bibinfo {year} {2000})}\BibitemShut {NoStop}%
\bibitem [{\citenamefont {Stark}(2001)}]{Stark:2001}%
  \BibitemOpen
  \bibfield  {author} {\bibinfo {author} {\bibfnamefont {H.}~\bibnamefont
  {Stark}},\ }\bibfield  {title} {\bibinfo {title} {Physics of colloidal
  dispersions in nematic liquid crystals},\ }\href
  {https://doi.org/https://doi.org/10.1016/S0370-1573(00)00144-7} {\bibfield
  {journal} {\bibinfo  {journal} {Phys. Rep.}\ }\textbf {\bibinfo {volume}
  {351}},\ \bibinfo {pages} {387 } (\bibinfo {year} {2001})}\BibitemShut
  {NoStop}%
\bibitem [{\citenamefont {Stark}\ and\ \citenamefont
  {Ventzki}(2001)}]{Stark:2001b}%
  \BibitemOpen
  \bibfield  {author} {\bibinfo {author} {\bibfnamefont {H.}~\bibnamefont
  {Stark}}\ and\ \bibinfo {author} {\bibfnamefont {D.}~\bibnamefont
  {Ventzki}},\ }\bibfield  {title} {\bibinfo {title} {Stokes drag of spherical
  particles in a nematic environment at low {E}ricksen numbers},\ }\href
  {https://doi.org/10.1103/PhysRevE.64.031711} {\bibfield  {journal} {\bibinfo
  {journal} {Phys. Rev. E}\ }\textbf {\bibinfo {volume} {64}},\ \bibinfo
  {pages} {031711} (\bibinfo {year} {2001})}\BibitemShut {NoStop}%
\bibitem [{\citenamefont {Everts}(2020)}]{Everts:2020b}%
  \BibitemOpen
  \bibfield  {author} {\bibinfo {author} {\bibfnamefont {J.~C.}\ \bibnamefont
  {Everts}},\ }\bibfield  {title} {\bibinfo {title} {Screened coulomb
  interactions of general macroions with nonzero particle volume},\ }\href
  {https://doi.org/10.1103/PhysRevResearch.2.033144} {\bibfield  {journal}
  {\bibinfo  {journal} {Phys. Rev. Research}\ }\textbf {\bibinfo {volume}
  {2}},\ \bibinfo {pages} {033144} (\bibinfo {year} {2020})}\BibitemShut
  {NoStop}%
\bibitem [{\citenamefont {Trizac}\ \emph {et~al.}(2003)\citenamefont {Trizac},
  \citenamefont {Bocquet}, \citenamefont {Aubouy},\ and\ \citenamefont {von
  Grünberg}}]{Trizac:2003}%
  \BibitemOpen
  \bibfield  {author} {\bibinfo {author} {\bibfnamefont {E.}~\bibnamefont
  {Trizac}}, \bibinfo {author} {\bibfnamefont {L.}~\bibnamefont {Bocquet}},
  \bibinfo {author} {\bibfnamefont {M.}~\bibnamefont {Aubouy}},\ and\ \bibinfo
  {author} {\bibfnamefont {H.~H.}\ \bibnamefont {von Grünberg}},\ }\bibfield
  {title} {\bibinfo {title} {Alexander's prescription for colloidal charge
  renormalization},\ }\href {https://doi.org/10.1021/la027056m} {\bibfield
  {journal} {\bibinfo  {journal} {Langmuir}\ }\textbf {\bibinfo {volume}
  {19}},\ \bibinfo {pages} {4027} (\bibinfo {year} {2003})}\BibitemShut
  {NoStop}%
\bibitem [{\citenamefont {\ifmmode~\check{S}\else \v{S}\fi{}arlah}\ and\
  \citenamefont {\ifmmode~\check{Z}\else \v{Z}\fi{}umer}(2001)}]{Sarlah:2001}%
  \BibitemOpen
  \bibfield  {author} {\bibinfo {author} {\bibfnamefont {A.}~\bibnamefont
  {\ifmmode~\check{S}\else \v{S}\fi{}arlah}}\ and\ \bibinfo {author}
  {\bibfnamefont {S.}~\bibnamefont {\ifmmode~\check{Z}\else \v{Z}\fi{}umer}},\
  }\bibfield  {title} {\bibinfo {title} {Van der {W}aals interaction mediated
  by an optically uniaxial layer},\ }\href
  {https://doi.org/10.1103/PhysRevE.64.051606} {\bibfield  {journal} {\bibinfo
  {journal} {Phys. Rev. E}\ }\textbf {\bibinfo {volume} {64}},\ \bibinfo
  {pages} {051606} (\bibinfo {year} {2001})}\BibitemShut {NoStop}%
\bibitem [{\citenamefont {Sonin}(1987)}]{Sonin:1987}%
  \BibitemOpen
  \bibfield  {author} {\bibinfo {author} {\bibfnamefont {A.~S.}\ \bibnamefont
  {Sonin}},\ }\bibfield  {title} {\bibinfo {title} {Lyotropic nematics},\
  }\href {https://doi.org/10.1070/pu1987v030n10abeh002967} {\bibfield
  {journal} {\bibinfo  {journal} {Sovi. Phys. Uspekhi}\ }\textbf {\bibinfo
  {volume} {30}},\ \bibinfo {pages} {875} (\bibinfo {year} {1987})}\BibitemShut
  {NoStop}%
\bibitem [{\citenamefont {Drwenski}\ \emph {et~al.}(2016)\citenamefont
  {Drwenski}, \citenamefont {Dussi}, \citenamefont {Hermes}, \citenamefont
  {Dijkstra},\ and\ \citenamefont {van Roij}}]{Drwenski:2016}%
  \BibitemOpen
  \bibfield  {author} {\bibinfo {author} {\bibfnamefont {T.}~\bibnamefont
  {Drwenski}}, \bibinfo {author} {\bibfnamefont {S.}~\bibnamefont {Dussi}},
  \bibinfo {author} {\bibfnamefont {M.}~\bibnamefont {Hermes}}, \bibinfo
  {author} {\bibfnamefont {M.}~\bibnamefont {Dijkstra}},\ and\ \bibinfo
  {author} {\bibfnamefont {R.}~\bibnamefont {van Roij}},\ }\bibfield  {title}
  {\bibinfo {title} {Phase diagrams of charged colloidal rods: Can a uniaxial
  charge distribution break chiral symmetry?},\ }\href
  {https://doi.org/10.1063/1.4942772} {\bibfield  {journal} {\bibinfo
  {journal} {J. Chem. Phys.}\ }\textbf {\bibinfo {volume} {144}},\ \bibinfo
  {pages} {094901} (\bibinfo {year} {2016})}\BibitemShut {NoStop}%
\bibitem [{\citenamefont {Takeuchi}\ and\ \citenamefont
  {Sano}(2010)}]{Takeuchi:2010}%
  \BibitemOpen
  \bibfield  {author} {\bibinfo {author} {\bibfnamefont {K.~A.}\ \bibnamefont
  {Takeuchi}}\ and\ \bibinfo {author} {\bibfnamefont {M.}~\bibnamefont
  {Sano}},\ }\bibfield  {title} {\bibinfo {title} {Universal fluctuations of
  growing interfaces: Evidence in turbulent liquid crystals},\ }\href
  {https://doi.org/10.1103/PhysRevLett.104.230601} {\bibfield  {journal}
  {\bibinfo  {journal} {Phys. Rev. Lett.}\ }\textbf {\bibinfo {volume} {104}},\
  \bibinfo {pages} {230601} (\bibinfo {year} {2010})}\BibitemShut {NoStop}%
\bibitem [{\citenamefont {Shah}\ and\ \citenamefont
  {Abbott}(2001)}]{Shah:2001}%
  \BibitemOpen
  \bibfield  {author} {\bibinfo {author} {\bibfnamefont {R.~R.}\ \bibnamefont
  {Shah}}\ and\ \bibinfo {author} {\bibfnamefont {N.~L.}\ \bibnamefont
  {Abbott}},\ }\bibfield  {title} {\bibinfo {title} {Coupling of the
  orientations of liquid crystals to electrical double layers formed by the
  dissociation of surface-immobilized salts},\ }\href
  {https://doi.org/10.1021/jp004073g} {\bibfield  {journal} {\bibinfo
  {journal} {J. Phys. Chem. B}\ }\textbf {\bibinfo {volume} {105}},\ \bibinfo
  {pages} {4936} (\bibinfo {year} {2001})}\BibitemShut {NoStop}%
\bibitem [{\citenamefont {Tojo}\ \emph {et~al.}(2009)\citenamefont {Tojo},
  \citenamefont {Furukawa}, \citenamefont {Araki},\ and\ \citenamefont
  {Onuki}}]{Onuki:2009}%
  \BibitemOpen
  \bibfield  {author} {\bibinfo {author} {\bibfnamefont {K.}~\bibnamefont
  {Tojo}}, \bibinfo {author} {\bibfnamefont {A.}~\bibnamefont {Furukawa}},
  \bibinfo {author} {\bibfnamefont {T.}~\bibnamefont {Araki}},\ and\ \bibinfo
  {author} {\bibfnamefont {A.}~\bibnamefont {Onuki}},\ }\bibfield  {title}
  {\bibinfo {title} {Defect structures in nematic liquid crystals around
  charged particles},\ }\href {https://doi.org/10.1140/epje/i2009-10506-7}
  {\bibfield  {journal} {\bibinfo  {journal} {Eur. Phys. J. E}\ }\textbf
  {\bibinfo {volume} {30}},\ \bibinfo {pages} {55} (\bibinfo {year}
  {2009})}\BibitemShut {NoStop}%
\bibitem [{\citenamefont {Everts}\ and\ \citenamefont
  {Ravnik}(2020)}]{Everts:2020}%
  \BibitemOpen
  \bibfield  {author} {\bibinfo {author} {\bibfnamefont {J.~C.}\ \bibnamefont
  {Everts}}\ and\ \bibinfo {author} {\bibfnamefont {M.}~\bibnamefont
  {Ravnik}},\ }\bibfield  {title} {\bibinfo {title} {Charge-, salt- and
  flexoelectricity-driven anchoring effects in nematics},\ }\bibfield
  {journal} {\bibinfo  {journal} {Liq. Cryst.}\ }\href
  {https://doi.org/10.1080/02678292.2020.1786176}
  {10.1080/02678292.2020.1786176} (\bibinfo {year} {2020})\BibitemShut
  {NoStop}%
\bibitem [{\citenamefont {Ravnik}\ and\ \citenamefont
  {Everts}(2020)}]{Ravnik:2020}%
  \BibitemOpen
  \bibfield  {author} {\bibinfo {author} {\bibfnamefont {M.}~\bibnamefont
  {Ravnik}}\ and\ \bibinfo {author} {\bibfnamefont {J.~C.}\ \bibnamefont
  {Everts}},\ }\bibfield  {title} {\bibinfo {title} {Topological-defect-induced
  surface charge heterogeneities in nematic electrolytes},\ }\href
  {https://doi.org/10.1103/PhysRevLett.125.037801} {\bibfield  {journal}
  {\bibinfo  {journal} {Phys. Rev. Lett.}\ }\textbf {\bibinfo {volume} {125}},\
  \bibinfo {pages} {037801} (\bibinfo {year} {2020})}\BibitemShut {NoStop}%
\bibitem [{\citenamefont {Paladugu}\ \emph {et~al.}(2017)\citenamefont
  {Paladugu}, \citenamefont {Conklin}, \citenamefont {Vi\~nals},\ and\
  \citenamefont {Lavrentovich}}]{Lavrentovich:2017}%
  \BibitemOpen
  \bibfield  {author} {\bibinfo {author} {\bibfnamefont {S.}~\bibnamefont
  {Paladugu}}, \bibinfo {author} {\bibfnamefont {C.}~\bibnamefont {Conklin}},
  \bibinfo {author} {\bibfnamefont {J.}~\bibnamefont {Vi\~nals}},\ and\
  \bibinfo {author} {\bibfnamefont {O.~D.}\ \bibnamefont {Lavrentovich}},\
  }\bibfield  {title} {\bibinfo {title} {Nonlinear electrophoresis of colloids
  controlled by anisotropic conductivity and permittivity of liquid-crystalline
  electrolyte},\ }\href {https://doi.org/10.1103/PhysRevApplied.7.034033}
  {\bibfield  {journal} {\bibinfo  {journal} {Phys. Rev. Applied}\ }\textbf
  {\bibinfo {volume} {7}},\ \bibinfo {pages} {034033} (\bibinfo {year}
  {2017})}\BibitemShut {NoStop}%
\bibitem [{\citenamefont {B.~Wu}\ and\ \citenamefont {Zhang}(2014)}]{Wu:2014}%
  \BibitemOpen
  \bibfield  {author} {\bibinfo {author} {\bibfnamefont {J.~C.}\ \bibnamefont
  {B.~Wu}, \bibfnamefont {Ye~Liu}}\ and\ \bibinfo {author} {\bibfnamefont
  {S.}~\bibnamefont {Zhang}},\ }\bibfield  {title} {\bibinfo {title} {Ligand
  dynamic effect on phase and morphology control of hexagonal nayf4},\ }\href
  {https://doi.org/10.1039/C4CE00109E} {\bibfield  {journal} {\bibinfo
  {journal} {Cryst. Eng. Comm.}\ }\textbf {\bibinfo {volume} {16}},\ \bibinfo
  {pages} {4472} (\bibinfo {year} {2014})}\BibitemShut {NoStop}%
\end{thebibliography}%
\mbox{}\\
{\bf Acknowledgments:} J.C.E. acknowledges fruitful discussions with S. \v Copar and A. Šarlah. {\bf Funding:} \mbox{J. C. E.} acknowledges financial support from the European Union's Horizon 2020 programme under the Marie Skłodowska-Curie grant agreement no. 795377 and from the Polish National Agency for Academic Exchange (NAWA) under the Ulam programme grant no. PPN/ULM/2019/1/00257. M.R. acknowledges financial support from the Slovenian Research Agency ARRS under contracts P1-0099, J1-1697, and L1-8135. B.S., H.M. and I.I.S. acknowledge funding from the US Department of Energy, Office of Basic Energy Sciences, Division of Materials Sciences and Engineering, under award ER46921, contract DE-SC0019293 with the University of Colorado at Boulder. Last, the authors would like to thank the Isaac Newton Institute for Mathematical Sciences for support and hospitality during the program [The Mathematical Design of New Materials] when work on this paper was undertaken. I.I.S. also acknowledges the hospitality of the Kavli Institute for Theoretical Physics at the University of California, Santa Barbara, where he was working on this publication during his extended stay and where his research was supported in part by the U.S. NSF under grant no. NSF PHY-1748958. This work was supported by EPSRC grant number EP/R014604/1. {\bf Author contributions:}
J.C.E. performed the numerical and analytical theoretical calculations under the supervision of M.R.; B.S. and H.M. performed experiments and analyzed experimental data under the supervision of I.I.S. All authors contributed to writing and discussing the manuscript. {\bf Competing interests:}
The authors declare that they have no competing interests. {\bf Data and materials availability:} All data needed to evaluate the conclusions in the paper are present in the paper and/or the Supplementary Materials. Additional data related to this paper may be requested from the authors.

\end{document}